\documentclass[article, twocolumn,
   aps, pra,
  amsmath,amssymb,
  longbibliography,
  ]{revtex4-2}
\usepackage{graphicx,color}
\usepackage{comment}
\usepackage{amsmath}
\usepackage{natbib}
\usepackage{epsfig}
\usepackage{comment}
\begin{document}

\title{Excess entropy scaling of the transverse sound speed in simple fluids}

\author{S. A. Khrapak}
\email{Email: Sergey.Khrapak@gmx.de} 

\affiliation{Joint Institute for High Temperatures, Russian Academy of Sciences, 125412 Moscow, Russia}

\begin{abstract}
A calculation of the transverse sound velocity as a function of excess entropy is presented for several simple fluids, including the Lennard-Jones, Yukawa, one-component plasma, inverse-power law (soft sphere) and hard sphere models. A quasi-universal character of this dependence is established, extending Rosenfeld's excess-entropy scaling of transport coefficients to the transverse sound velocity. The results are discussed in terms of the soft- to hard-sphere crossover and the Frenkel crossover between gas-like and liquid-like dynamics.
\end{abstract}

\date{\today}

\maketitle

\section{Introduction}

In 1977 Rosenfeld suggested a relation between the transport coefficients and the excess entropy of simple fluids~\cite{RosenfeldPRA1977}. Originally, he considered macroscopically reduced (expressed in units of interatomic separation, atomic mass and thermal velocity) self-diffusion and shear viscosity coefficients of the hard-sphere, soft-sphere, and one-component plasma fluids and demonstrated their quasi-universal dependence on the excess entropy. In addition, he showed that this quasi-universal entropy dependence is only slightly affected by the addition of an attractive potential in the case of the Lennard-Jones fluid. Approximately two decades later, Dzugutov proposed a quasi-universal scaling law for atomic self-diffusion, where a self-diffusion coefficient expressed in units of the effective hard-sphere diameter and Enskog collision frequency, is related to the excess entropy in the two-particle approximation~\cite{DzugutovNature1996}. Around the same time, Rosenfeld revisited and extended the original idea to include the thermal conductivity coefficient and considered the dilute gas regime~\cite{RosenfeldJPCM1999}. He also elaborated in detail on the application of the excess entropy scaling to Yukawa fluids and introduced the freezing temperature scaling of transport coefficients~\cite{RosenfeldPRE2000,RosenfeldJPCM2001}. Applications to molecular, confined, supercooled, and metallic liquids, as well as to mixtures, broadened the scope of excess-entropy scaling and helped identify cases when it breaks down; see the excellent overview by Dyre~\cite{DyreJCP2018}.

Originally, excess entropy scaling was considered as a semi-quantitative approximate relation between dynamics and thermodynamics rather than a theory~\cite{RosenfeldJPCM1999}. The isomorph theory of Roskilde-simple liquids~\cite{GnanJCP2009,DyreJPCB2014,SchroderJCP2014} provided a general framework for excess entropy scaling and offered insight into why it works so well in some systems but breaks down in others~\cite{DyreJCP2018}. Roskilde-simple liquids have isomorphs, which are curves of constant excess entropy along which some properly reduced structure and dynamics quantities are invariant to a good approximation. Reduced transport properties, such as diffusion, shear viscosity, and thermal conductivity coefficients, are approximate invariants. This explains why these reduced transport coefficients exhibit excess entropy scaling.

The isomorph theory implies that the excess entropy scaling is not limited to the reduced transport coefficients. All other reduced quantities that are isomorph invariants should also be expected to exhibit excess entropy scaling. Recently, it was shown that the transverse
sound velocity, Maxwell relaxation time, and shear modulus are isomorph invariants in the Lennard-Jones system~\cite{KnudsenPRE2021}. Similarly, it was demonstrated that the reduced transverse sound speed is an isomorph invariant in Yukawa fluids~\cite{YuPRE2024}. The ratio of the transverse sound velocity to the thermal velocity has been used to identify the location of the Frenkel line in Yukawa systems~\cite{HuangPRR2023,YuPRE2024,XuPRR2026}. 

The instantaneous transverse sound velocity (defined through the corresponding shear modulus, see below) is an important liquid property that influences the Maxwell (shear) relaxation time~\cite{MountainJCP1966,KhrapakPRE11_2024}, diffusion in the vibrational paradigm of dense-liquid dynamics~\cite{KhrapakMolecules12_2021,KhrapakPhysRep2024}, the Stokes–Einstein relation between self-diffusion and shear viscosity~\cite{ZwanzigJCP1983,KhrapakMolPhys2019,KhrapakPRE10_2021}, and the onset of the glass transition within the framework of the shoving model~\cite{DyreRMP2006,DyreJCP2012,KhrapakJCP2024_glass} among others. In addition, it can be measured experimentally, e.g. in liquid metals~\cite{HosokawaPRL2009,HosokawaJPCM2013,HosokawaJPCM2015} and dusty plasma fluids~\cite{PramanikPRL2002,NunomuraPRL2005,BandyopadhuayPLA2008}, and determined in computer simulations~\cite{OhtaPRL2000,GoreePRE2012,YangPRL2017,BrykJCP2017,KryuchkovSciRep2019,KhrapakJCP2019,KryuchkovJCP2021}. Thus, the potential quasi-universality of the transverse sound velocity with respect to excess entropy warrants careful examination.
The purpose of the present work is to examine this scaling across several simple model fluids. Among the model fluids considered are the Lennard-Jones, Yukawa, one-component plasma, inverse power law (soft-sphere), and hard sphere models. Below we briefly explain how the excess entropy and the transverse sound velocities have been calculated for these different models. We then present and discuss the main observations, focusing on the quasi-universality of the excess entropy scaling. The universality of transverse sound velocity based criterion of the Frenkel line is addressed as well.

\section{Calculation}

In the following, we consider a wide range of pairwise interactions, including purely repulsive interactions and those having long-range attractive branches. It is convenient to present the potentials in the general form
$\phi(r) = \epsilon f(x)$, where $x = r/\sigma$ is the distance divided by the characteristic length-scale of the potential and $\epsilon$ is the energy scale. In this notation, we have $f(x)=x^{-n}$ for soft spheres (SS)~\cite{HooverJCP1970,HooverJCP1971}; in this study we consider two special cases $n=6$ and $n=12$. The limit $n\rightarrow \infty$ corresponds to the fluid of hard spheres (HS). The case $n=1$ with a fixed neutralizing background added corresponds to the one-component plasma (OCP) fluid. For the Yukawa potential (also known as Debye-H\"uckel or screened Coulomb) $f(x)=\exp(-\kappa x)/x$, where $\kappa$ is the screening parameter. This potential is widely used in the context of colloidal suspensions and plasma-related systems, such as complex and dusty plasmas~\cite{TsytovichUFN1997,FortovUFN,FortovPR,ShuklaRMP2009,MorfillRMP2009,FortovBook,IvlevBook}. One of the most widely used models that captures much of the essential physics of simple atomic liquids is the $12-6$ Lennard-Jones (LJ) potential~\cite{LennardJones1924}, $f(x)=4(x^{-12}-x^{-6})$. 

The high-frequency (instantaneous) elastic shear modulus of a simple fluid can be expressed in terms of the
pair-interaction potential $\phi(r)$ and the radial distribution function (RDF) $g(r)$. A detailed derivation was provided by Zwanzig and Mountain~\cite{ZwanzigJCP1965} and the result is 
\begin{equation}\label{G_}
G_{\infty}=\rho T+\frac{2\pi \rho^2}{15}\int_0^{\infty}dr r^3 g(r)\left[r \phi''(r)+4\phi'(r)\right].
\end{equation}
where $\rho$ is the number density, $T$ is the temperature expressed in energy units ($=k_{\rm B}T$), and the primes denote derivatives with respect to distance. The instantaneous transverse sound velocity $c_t$ is defined using the instantaneous shear modulus as
\begin{equation}\label{ct}
G_{\infty}=\rho m c_t^2,
\end{equation}
where $m$ is the atomic mass. In the following, we refer to the transverse sound velocity only in the sense of Eq.~(\ref{ct}). The peculiarities of the transverse mode dispersion relation in the liquid state (absence of a propagating mode at long wavelengths known as the $k$ -gap~\cite{BalucaniBook,BolmatovPCL2015,GoreePRE2012,YangPRL2017,BrykJCP2017,OhtaPRL2000,KhrapakJCP2019,KryuchkovSciRep2019,TrachenkoBook}) are not discussed. 

For a given pairwise interaction potential $\phi(r)$, the RDF needed to calculate $G_{\infty}$ or $c_t$ can be obtained through numerical simulations (molecular dynamics or Monte Carlo) or from the integral equation of state theory (Ornstein–Zernike equation supplemented by an appropriate closure) for each state point. However, for some simple fluids like SS, OCP and LJ fluids, such a brute force approach is not necessary because the integrals in Eq.~(\ref{G_}) can be easily related to the known thermodynamic properties of the system. For other fluids they can be evaluated using other simple means.

Let us divide the energy $U$ and pressure $P$ of a liquid into ideal gas and excess (configurational) contributions and express these in units of $NT$ and $\rho T$, respectively. we get $U/NT=3/2+u_{\rm ex}$ and $P/\rho T=1+p_{\rm ex}$, where $N$ is the number of atoms, $u_{\rm ex}$ is the reduced excess energy and $p_{\rm ex}$ is the reduced excess pressure. The excess energy  $u_{\rm ex}$ can be expressed using $g(r)$ and $\phi(r)$ from the energy equation~\cite{HansenBook}
\begin{equation}\label{energy}
u_{\rm ex} = \frac{2\pi\rho}{T}\int_0^{\infty}r^2\phi(r)g(r)dr.    
\end{equation}
Similarly  $p_{\rm ex}$ can be expressed using $g(r)$ and $\phi'(r)$ from the pressure equation (virial theorem)~\cite{HansenBook}
\begin{equation}\label{pressure}
p_{\rm ex} = -\frac{2\pi\rho}{3T}\int_0^{\infty}r^3\phi'(r)g(r)dr.    
\end{equation}
The excess entropy per particle in units of $k_{\rm B}$ is introduced through $s_{\rm ex}=s-s_{\rm id}$, where $s_{\rm id}$ is the reduced entropy of an ideal gas at the same temperature and density. 

For the SS fluid, combining Eqs.~(\ref{G_}) and (\ref{pressure}), we immediately get~\cite{HeyesJCP1994,KhrapakSciRep2017}
\begin{equation}\label{ct_SS}
 \frac{c_t^2}{v_{\rm T}^2}=1+\frac{n-3}{5}p_{\rm ex},  \end{equation}
where $v_{\rm T}=\sqrt{T/m}$ is the thermal velocity.
The parameters $p_{\rm ex}$ and $s_{\rm ex}$ are then inferred from the fits provided recently in Ref.~\cite{HeyesJCP2025}. Note that Eq.~(\ref{ct_SS}) remains formally valid in the OCP limit ($n=1$), where the presence of a neutralizing background makes the internal pressure and energy finite and negative~\cite{BausPR1980,GoldenPoP2000,KhrapakPoP2016}. The excess pressure and entropy of the OCP are calculated here from the fits for the excess energy and the excess Helmholtz free energy suggested in Red.~\cite{KhrapakCPP2016}. 

For the LJ system, the transverse sound velocity can be related to excess energy and pressure by expressing Eq.~(\ref{G_}) as a linear combination of Eqs.~(\ref{energy}) and (\ref{pressure}) taking into account (\ref{ct}). This leads to~\cite{ZwanzigJCP1965,KhrapakMolecules2020,KhrapakPRE12_2023}  
\begin{equation}\label{ct_LJ}
 \frac{c_t^2}{v_{\rm T}^2}=1-\frac{24}{5}u_{\rm ex}+3p_{\rm ex}.   
\end{equation}
In this case, all thermodynamic quantities, including the excess entropy, are calculated from the equation of state by Thol {\it et al}.~\cite{Thol2016}.

In the HS limit ($n\rightarrow\infty$), Eq.~(\ref{ct_SS}) predicts divergence of the transverse sound velocity and shear modulus because $p_{\rm ex}$ remains finite in this limit~\cite{HeyesJCP1994,Frisch1966}. This problem has recently been addressed~\cite{KhrapakSciRep2017,KhrapakPRE09_2019} and it was concluded that the divergence is artificial, implying that Eq.~(\ref{ct_SS}) is not appropriate in this limit~\cite{KhrapakPRE05_2021}. Consequently, we use the HS shear modulus as derived by Miller~\cite{MillerJCP1969} and discussed by Khrapak~\cite{KhrapakPRE09_2019}. Miller first derived a general expression for the stress tensor of a monatomic system of rigid spheres. He then obtained the elastic moduli by specializing Green's method to a hard-sphere fluid under the conditions of local equilibrium~\cite{MillerJCP1969}. His result for the instantaneous shear modulus, expressed using the present notation, reads~\cite{KhrapakPRE09_2019}
\begin{equation}
 G_{\infty}=\rho T\left[1-\frac85\eta g'(\sigma)\right],   
\end{equation}
where $\eta=\pi\rho\sigma^3/6$ is the packing fraction, $\sigma$ is the diameter of the hard sphere and $g'(\sigma)$ is the derivative of the RDF at contact. Here we use an approximation for  $g'(\sigma)$ proposed by Tao {\it et al}.~\cite{TaoPRA1992}. The excess entropy of the HS fluid is calculated from the Carnahan-Starling equation of state~\cite{CarnahanJCP1969}. 

For the Yukawa system, no direct relation between the elastic moduli and the thermodynamic properties is also available. However, there is an elegant method that provides access to both -- the variational approach based on the Bogoliubov inequality with HS fluid as a reference system~\cite{OCP_Variational,YukawaSubmitted}. For this calculation, we use the variational approach to estimate the instantaneous shear modulus and the transverse sound velocity; 
%the transverse sound velocity is estimated using the Percus-Yevick RDF of the HS fluid, evaluated at an unphysical packing fraction $\eta=1$. This usually provides a rather good estimate for the thermodynamic properties and elastic moduli of a strongly coupled Yukawa fluid~\cite{YukawaSubmitted,KhrapakPoP10_2019}. The expression used is 
%\begin{equation}
% \frac{c_t^2}{v_{\rm T}^2} = 1+\frac{\Gamma\kappa^4\left[(3+\kappa^2)\sinh(\kappa)-3\kappa\cosh(\kappa)\right]}{15\left[\kappa\cosh(\kappa)-\sinh(\kappa)\right]^3}. 
%\end{equation}
the excess entropy is obtained from the fits of Ref.~\cite{KhrapakPRE09_2024}, based on extensive results from MD simulations~\cite{HamaguchiPRE1997}.     

\begin{figure*}
\includegraphics[width=15cm]{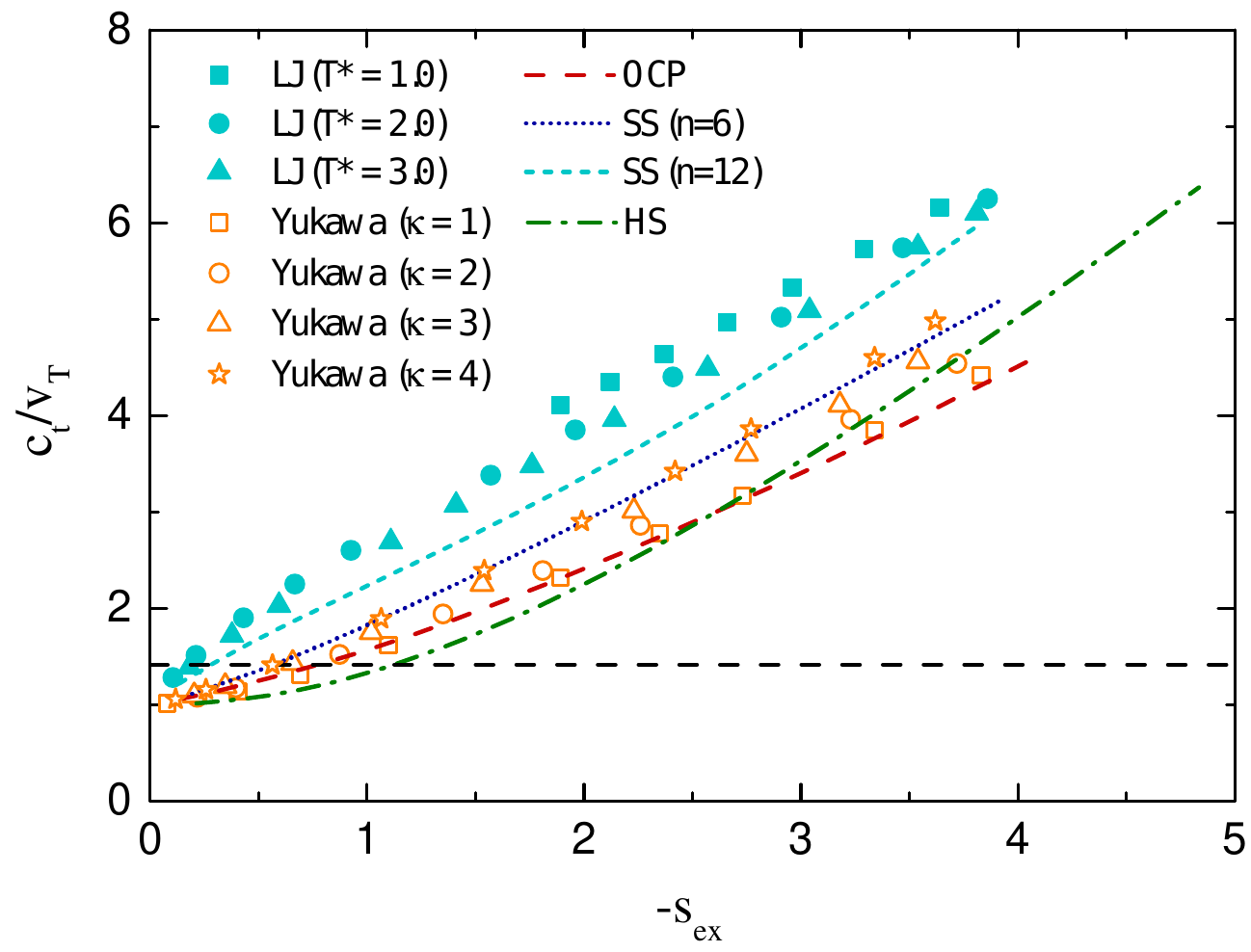}
\caption{(Color online) Reduced transverse sound speed $c_t/v_{\rm T}$ versus the negative value of the excess entropy $-s_{\rm ex}$ for various simple fluids. Solid symbols correspond to the Lennard-Jones fluid along three isotherms (see the legend). Open symbols correspond to the Yukawa fluid with different screening parameters (see the legend for details). The dashed line corresponds to the OCP fluid. The short-dotted and the dotted curves are the SS results for $n=6$ and $n=12$, respectively. The dash-dotted curve corresponds to the HS fluid.}
\label{Fig1}
\end{figure*}

\section{Results}

The results of this calculation are shown in Fig.~\ref{Fig1}. The quasi-universal excess-entropy scaling of the transverse sound velocity is established. The quality of this quasi-universality seems to be comparable to that of the excess entropy scaling of transport coefficients demonstrated by Rosenfeld~\cite{RosenfeldPRA1977}. The quasi-universality lies in the fact that when the transverse sound velocity is reduced by the thermal velocity, its value can be estimated within about $30 -50\%$ based on the excess entropy value. At the same time, the nominal transverse sound velocities can vary greatly from one physical system to another. For example, the typical value of $c_t$ in liquid metals is on the order of $10^5$ cm/s~\cite{HosokawaJPCM2015,MorkelPRE1993}. The same order of magnitude has been reported for water~\cite{RahmanPRA1974,PontecorvoPRE2005}. In complex (dusty) plasmas --  classical systems of highly charged micron-size particles immersed in a plasma medium -- typical measured transverse sound velocities are five orders of magnitude lower, $c_t\sim 1$ cm/s~\cite{NosenkoPRL2002,NosenkoPRE2003,BandyopadhuayPLA2008}.

The main expected key reference points are well reproduced in Fig.~\ref{Fig1}. For example, the fluid-solid phase transition for soft potentials is located at $s_{\rm ex}\simeq -4$~\cite{RosenfeldPRE2000,HeyesJCP2025,KhrapakJCP2022_1} and at $s_{\rm ex}\simeq -4.8$ in the HS limit~\cite{HeyesJCP2025}. The corresponding transverse sound velocities near the fluid-solid phase transition are $c_t/v_{\rm T}\simeq 6$ for the LJ system~\cite{KhrapakMolecules2020}, $c_t/v_{\rm T}\simeq 5 - 6$ for the SS system with $n<12$~\cite{KhrapakMolecules2020,KhrapakPoF2023}, $c_t/v_{\rm T}\simeq 5$ for the Yukawa system~\cite{KhrapakPoF2023,YuPRE2024}, $c_t/v_{\rm T}\simeq 4.6$ for the OCP~\cite{KhrapakPRR2020}, and $c_t/v_{\rm T}\simeq 6.4$ in the HS limit~\cite{KhrapakPoF2023}.    

According to Fig.~\ref{Fig1}, a nearly linear dependence of  $c_t/v_{\rm T}$ on $-s_{\rm ex}$ is evident, with the slope increasing systematically as the interaction potential becomes steeper from the OCP limit to the Yukawa potential with increasing $\kappa$, followed by the soft-sphere model with increasing $n$, and ultimately the Lennard–Jones model. A drop is then observed for the HS limiting curve, with the transverse sound velocity in the HS fluid becoming comparable to, or even lower than, that in the OCP fluid. This non-monotonic behavior is not particularly surprising or unexpected. However, an important remark regarding the crossover between soft-sphere and hard-sphere fluids is in order. In Ref.~\cite{KhrapakPRE05_2021}, the excitation spectra of the inverse-power-law (IPL) fluids were analyzed in detail and the corresponding elastic sound velocities were determined from the dispersion relations. It was observed that the classical Zwanzig-Mountain formula (\ref{G_}) can overestimate the measured transverse sound velocities at large IPL (SS) exponents $n$. In practice, Eq.~(\ref{G_}) seems to already have become unreliable at $n\gtrsim 10$~\cite{KhrapakPRE05_2021}. This may affect the SS ($n=12$) results, but even more the LJ results depicted in Fig.~\ref{Fig1}.  It might seem somewhat surprising, but the effective IPL exponent for the LJ potential can considerably exceed that of its repulsive $\propto r^{-12}$ part. In particular, at moderate densities $n_{\rm eff}\simeq 18$~\cite{PedersenPRL2008}. Thus, the values of $c_t/v_{\rm T}$ predicted for the LJ fluid with the help of Eq.~(\ref{G_}) may be overestimated. This issue deserves a careful examination in the future. If confirmed, it would imply an even better quasi-universality of excess-entropy scaling. Since we know that Zwanzig-Mountain expressions tend to overestimate the elastic moduli of steep potentials, the results presented here for the SS and LJ models can be considered as upper limits on their actual values.   

The horizontal dashed line in Fig.~\ref{Fig1} corresponds to the condition $c_t/v_{\rm T}= \sqrt{2}$ and has a clear physical meaning: it corresponds to the point where the ideal gas (kinetic) and configurational (excess) contributions to the instantaneous shear modulus and transverse sound velocity are equal. Therefore, it was proposed that this point may serve as an appropriate criterion to mark the crossover between gas-like and liquid-like dynamics~\cite{HuangPRR2023,YuPRE2024,XuPRR2026}, known as the Frenkel crossover~\cite{BrazhkinPRE2012,BrazhkinPRL2013,BrazhkinUFN2012,KhrapakPRE07_2025}. Gas-like dynamics occurs in regimes characterized by dominance of the kinetic contribution, whereas liquid-like dynamics arises when the configurational contribution prevails. In another approach to the Frenkel line, excess entropy has been suggested as an appropriate criterion to locate the crossover~\cite{BellJCP2020}. In particular, condition $s_{\rm ex}\simeq -1$ can serve as a quasi-universal indication of the Frenkel line~\cite{BellJCP2020,KhrapakJCP2022,KhrapakPRE10_2021,KhrapakPRE07_2025}. We see in Fig.~~\ref{Fig1} that for sufficiently soft potentials (as well as in the HS limit) the conditions $c_t/v_{\rm T}= \sqrt{2}$ and $s_{\rm ex}\simeq -1$ agree with reasonable accuracy. However, as the steepness of the Yukawa and SS potentials increases, the intersection moves to less negative values of the excess entropy. This tendency is particularly pronounced for the SS ($n = 12$) and LJ potentials. However, as discussed above, this unexpected trend may result from an overestimation of the ratio $c_t / v_{\rm T}$ arising from the use of Zwanzig-Mountain Eq.~(\ref{G_}). This likely warrants further investigation.  

\section{Conclusion}

An approximately quasi-universal dependence of the reduced transverse sound velocity on the excess entropy has been demonstrated here for various simple fluids with disparate pairwise interactions. This includes the well known Lennard-Jones, Yukawa, one-component plasma, inverse-power law (soft sphere) and hard sphere models. The dependence of the reduced transverse sound velocity on  $-s_{\rm ex}$ is almost linear, with the slope increasing systematically as the interaction potential becomes steeper. However, the Zwanzig–Mountain expression used to calculate the transverse sound velocity tends to overestimate its value as the potential steepness exceeds a certain threshold~\cite{KhrapakPRE05_2021}. This may imply that the quasi-universality of the scaling is even more pronounced than suggested by the present analysis. In the regime of sufficiently soft interactions, where no accuracy issue arises, the two criteria used to locate the Frenkel crossover, $c_t/v_{\rm T}= \sqrt{2}$ and  $s_{\rm ex}\simeq -1$, show reasonable agreement. The discussed approximate quasi-universality is expected to be more pronounced for densities above and excess entropy below those at the Frenkel crossover. In this regime, many liquids exhibit similar behavior due to the dominant contribution of vibrational motion in their thermodynamic and transport properties~\cite{KhrapakPhysRep2024}.

%The author declares no conflict of interests.

%The data that support the findings of this study are available from the authors upon reasonable request. 

\bibliography{SE_Ref}

%apsrev4-2.bst 2019-01-14 (MD) hand-edited version of apsrev4-1.bst
%Control: key (0)
%Control: author (8) initials jnrlst
%Control: editor formatted (1) identically to author
%Control: production of article title (0) allowed
%Control: page (0) single
%Control: year (1) truncated
%Control: production of eprint (0) enabled
\providecommand{\noopsort}[1]{}\providecommand{\singleletter}[1]{#1}%
\begin{thebibliography}{86}%
\makeatletter
\providecommand \@ifxundefined [1]{%
 \@ifx{#1\undefined}
}%
\providecommand \@ifnum [1]{%
 \ifnum #1\expandafter \@firstoftwo
 \else \expandafter \@secondoftwo
 \fi
}%
\providecommand \@ifx [1]{%
 \ifx #1\expandafter \@firstoftwo
 \else \expandafter \@secondoftwo
 \fi
}%
\providecommand \natexlab [1]{#1}%
\providecommand \enquote  [1]{``#1''}%
\providecommand \bibnamefont  [1]{#1}%
\providecommand \bibfnamefont [1]{#1}%
\providecommand \citenamefont [1]{#1}%
\providecommand \href@noop [0]{\@secondoftwo}%
\providecommand \href [0]{\begingroup \@sanitize@url \@href}%
\providecommand \@href[1]{\@@startlink{#1}\@@href}%
\providecommand \@@href[1]{\endgroup#1\@@endlink}%
\providecommand \@sanitize@url [0]{\catcode `\\12\catcode `\$12\catcode
  `\&12\catcode `\#12\catcode `\^12\catcode `\_12\catcode `\%12\relax}%
\providecommand \@@startlink[1]{}%
\providecommand \@@endlink[0]{}%
\providecommand \url  [0]{\begingroup\@sanitize@url \@url }%
\providecommand \@url [1]{\endgroup\@href {#1}{\urlprefix }}%
\providecommand \urlprefix  [0]{URL }%
\providecommand \Eprint [0]{\href }%
\providecommand \doibase [0]{https://doi.org/}%
\providecommand \selectlanguage [0]{\@gobble}%
\providecommand \bibinfo  [0]{\@secondoftwo}%
\providecommand \bibfield  [0]{\@secondoftwo}%
\providecommand \translation [1]{[#1]}%
\providecommand \BibitemOpen [0]{}%
\providecommand \bibitemStop [0]{}%
\providecommand \bibitemNoStop [0]{.\EOS\space}%
\providecommand \EOS [0]{\spacefactor3000\relax}%
\providecommand \BibitemShut  [1]{\csname bibitem#1\endcsname}%
\let\auto@bib@innerbib\@empty
%</preamble>
\bibitem [{\citenamefont {Rosenfeld}(1977)}]{RosenfeldPRA1977}%
  \BibitemOpen
  \bibfield  {author} {\bibinfo {author} {\bibfnamefont {Y.}~\bibnamefont
  {Rosenfeld}},\ }\bibfield  {title} {\bibinfo {title} {Relation between the
  transport coefficients and the internal entropy of simple systems},\ }\href
  {https://doi.org/10.1103/physreva.15.2545} {\bibfield  {journal} {\bibinfo
  {journal} {Phys. Rev. A}\ }\textbf {\bibinfo {volume} {15}},\ \bibinfo
  {pages} {2545} (\bibinfo {year} {1977})}\BibitemShut {NoStop}%
\bibitem [{\citenamefont {Dzugutov}(1996)}]{DzugutovNature1996}%
  \BibitemOpen
  \bibfield  {author} {\bibinfo {author} {\bibfnamefont {M.}~\bibnamefont
  {Dzugutov}},\ }\bibfield  {title} {\bibinfo {title} {A universal scaling law
  for atomic diffusion in condensed matter},\ }\href
  {https://doi.org/10.1038/381137a0} {\bibfield  {journal} {\bibinfo  {journal}
  {Nature}\ }\textbf {\bibinfo {volume} {381}},\ \bibinfo {pages} {137}
  (\bibinfo {year} {1996})}\BibitemShut {NoStop}%
\bibitem [{\citenamefont {Rosenfeld}(1999)}]{RosenfeldJPCM1999}%
  \BibitemOpen
  \bibfield  {author} {\bibinfo {author} {\bibfnamefont {Y.}~\bibnamefont
  {Rosenfeld}},\ }\bibfield  {title} {\bibinfo {title} {A quasi-universal
  scaling law for atomic transport in simple fluids},\ }\href
  {https://doi.org/10.1088/0953-8984/11/28/303} {\bibfield  {journal} {\bibinfo
   {journal} {J. Phys.: Condens. Matter}\ }\textbf {\bibinfo {volume} {11}},\
  \bibinfo {pages} {5415} (\bibinfo {year} {1999})}\BibitemShut {NoStop}%
\bibitem [{\citenamefont {Rosenfeld}(2000)}]{RosenfeldPRE2000}%
  \BibitemOpen
  \bibfield  {author} {\bibinfo {author} {\bibfnamefont {Y.}~\bibnamefont
  {Rosenfeld}},\ }\bibfield  {title} {\bibinfo {title} {Excess-entropy and
  freezing-temperature scalings for transport coefficients: Self-diffusion in
  {Y}ukawa systems},\ }\href {https://doi.org/10.1103/physreve.62.7524}
  {\bibfield  {journal} {\bibinfo  {journal} {Phys. Rev. E}\ }\textbf {\bibinfo
  {volume} {62}},\ \bibinfo {pages} {7524} (\bibinfo {year}
  {2000})}\BibitemShut {NoStop}%
\bibitem [{\citenamefont {Rosenfeld}(2001)}]{RosenfeldJPCM2001}%
  \BibitemOpen
  \bibfield  {author} {\bibinfo {author} {\bibfnamefont {Y.}~\bibnamefont
  {Rosenfeld}},\ }\bibfield  {title} {\bibinfo {title} {Quasi-universal
  melting-temperature scaling of transport coefficients in {Y}ukawa systems},\
  }\href {https://doi.org/10.1088/0953-8984/13/2/101} {\bibfield  {journal}
  {\bibinfo  {journal} {J. Phys.: Condens. Matter}\ }\textbf {\bibinfo {volume}
  {13}},\ \bibinfo {pages} {L39} (\bibinfo {year} {2001})}\BibitemShut
  {NoStop}%
\bibitem [{\citenamefont {Dyre}(2018)}]{DyreJCP2018}%
  \BibitemOpen
  \bibfield  {author} {\bibinfo {author} {\bibfnamefont {J.~C.}\ \bibnamefont
  {Dyre}},\ }\bibfield  {title} {\bibinfo {title} {Perspective:
  {E}xcess-entropy scaling},\ }\href {https://doi.org/10.1063/1.5055064}
  {\bibfield  {journal} {\bibinfo  {journal} {J. Chem. Phys.}\ }\textbf
  {\bibinfo {volume} {149}},\ \bibinfo {pages} {210901} (\bibinfo {year}
  {2018})}\BibitemShut {NoStop}%
\bibitem [{\citenamefont {Gnan}\ \emph {et~al.}(2009)\citenamefont {Gnan},
  \citenamefont {Schroder}, \citenamefont {Pedersen}, \citenamefont {Bailey},\
  and\ \citenamefont {Dyre}}]{GnanJCP2009}%
  \BibitemOpen
  \bibfield  {author} {\bibinfo {author} {\bibfnamefont {N.}~\bibnamefont
  {Gnan}}, \bibinfo {author} {\bibfnamefont {T.~B.}\ \bibnamefont {Schroder}},
  \bibinfo {author} {\bibfnamefont {U.~R.}\ \bibnamefont {Pedersen}}, \bibinfo
  {author} {\bibfnamefont {N.~P.}\ \bibnamefont {Bailey}},\ and\ \bibinfo
  {author} {\bibfnamefont {J.~C.}\ \bibnamefont {Dyre}},\ }\bibfield  {title}
  {\bibinfo {title} {Pressure-energy correlations in liquids. {IV}. {I}somorphs
  in liquid phase diagrams},\ }\href {https://doi.org/10.1063/1.3265957}
  {\bibfield  {journal} {\bibinfo  {journal} {J. Chem. Phys.}\ }\textbf
  {\bibinfo {volume} {131}},\ \bibinfo {pages} {234504} (\bibinfo {year}
  {2009})}\BibitemShut {NoStop}%
\bibitem [{\citenamefont {Dyre}(2014)}]{DyreJPCB2014}%
  \BibitemOpen
  \bibfield  {author} {\bibinfo {author} {\bibfnamefont {J.~C.}\ \bibnamefont
  {Dyre}},\ }\bibfield  {title} {\bibinfo {title} {Hidden scale invariance in
  condensed matter},\ }\href {https://doi.org/10.1021/jp501852b} {\bibfield
  {journal} {\bibinfo  {journal} {J. Phys. Chem. B}\ }\textbf {\bibinfo
  {volume} {118}},\ \bibinfo {pages} {10007} (\bibinfo {year}
  {2014})}\BibitemShut {NoStop}%
\bibitem [{\citenamefont {Schr{\o}der}\ and\ \citenamefont
  {Dyre}(2014)}]{SchroderJCP2014}%
  \BibitemOpen
  \bibfield  {author} {\bibinfo {author} {\bibfnamefont {T.~B.}\ \bibnamefont
  {Schr{\o}der}}\ and\ \bibinfo {author} {\bibfnamefont {J.~C.}\ \bibnamefont
  {Dyre}},\ }\bibfield  {title} {\bibinfo {title} {Simplicity of condensed
  matter at its core: {G}eneric definition of a {R}oskilde-simple system},\
  }\href {https://doi.org/10.1063/1.4901215} {\bibfield  {journal} {\bibinfo
  {journal} {J. Chem. Phys.}\ }\textbf {\bibinfo {volume} {141}},\ \bibinfo
  {pages} {204502} (\bibinfo {year} {2014})}\BibitemShut {NoStop}%
\bibitem [{\citenamefont {Knudsen}\ \emph {et~al.}(2021)\citenamefont
  {Knudsen}, \citenamefont {Todd}, \citenamefont {Dyre},\ and\ \citenamefont
  {Hansen}}]{KnudsenPRE2021}%
  \BibitemOpen
  \bibfield  {author} {\bibinfo {author} {\bibfnamefont {S.}~\bibnamefont
  {Knudsen}}, \bibinfo {author} {\bibfnamefont {B.~D.}\ \bibnamefont {Todd}},
  \bibinfo {author} {\bibfnamefont {J.~C.}\ \bibnamefont {Dyre}},\ and\
  \bibinfo {author} {\bibfnamefont {J.~S.}\ \bibnamefont {Hansen}},\ }\bibfield
   {title} {\bibinfo {title} {Generalized hydrodynamics of the
  {L}ennard-{J}ones liquid in view of hidden scale invariance},\ }\href
  {https://doi.org/10.1103/physreve.104.054126} {\bibfield  {journal} {\bibinfo
   {journal} {Phys. Rev. E}\ }\textbf {\bibinfo {volume} {104}},\ \bibinfo
  {pages} {054126} (\bibinfo {year} {2021})}\BibitemShut {NoStop}%
\bibitem [{\citenamefont {Yu}\ \emph {et~al.}(2024)\citenamefont {Yu},
  \citenamefont {Huang}, \citenamefont {Lu}, \citenamefont {Khrapak},\ and\
  \citenamefont {Feng}}]{YuPRE2024}%
  \BibitemOpen
  \bibfield  {author} {\bibinfo {author} {\bibfnamefont {N.}~\bibnamefont
  {Yu}}, \bibinfo {author} {\bibfnamefont {D.}~\bibnamefont {Huang}}, \bibinfo
  {author} {\bibfnamefont {S.}~\bibnamefont {Lu}}, \bibinfo {author}
  {\bibfnamefont {S.}~\bibnamefont {Khrapak}},\ and\ \bibinfo {author}
  {\bibfnamefont {Y.}~\bibnamefont {Feng}},\ }\bibfield  {title} {\bibinfo
  {title} {Universal scaling of transverse sound speed and its isomorphic
  property in {Y}ukawa fluids},\ }\href
  {https://doi.org/10.1103/physreve.109.035202} {\bibfield  {journal} {\bibinfo
   {journal} {Phys. Rev. E}\ }\textbf {\bibinfo {volume} {109}},\ \bibinfo
  {pages} {035202} (\bibinfo {year} {2024})}\BibitemShut {NoStop}%
\bibitem [{\citenamefont {Huang}\ \emph {et~al.}(2023)\citenamefont {Huang},
  \citenamefont {Baggioli}, \citenamefont {Lu}, \citenamefont {Ma},\ and\
  \citenamefont {Feng}}]{HuangPRR2023}%
  \BibitemOpen
  \bibfield  {author} {\bibinfo {author} {\bibfnamefont {D.}~\bibnamefont
  {Huang}}, \bibinfo {author} {\bibfnamefont {M.}~\bibnamefont {Baggioli}},
  \bibinfo {author} {\bibfnamefont {S.}~\bibnamefont {Lu}}, \bibinfo {author}
  {\bibfnamefont {Z.}~\bibnamefont {Ma}},\ and\ \bibinfo {author}
  {\bibfnamefont {Y.}~\bibnamefont {Feng}},\ }\bibfield  {title} {\bibinfo
  {title} {Revealing the supercritical dynamics of dusty plasmas and their
  liquidlike to gaslike dynamical crossover},\ }\href
  {https://doi.org/10.1103/physrevresearch.5.013149} {\bibfield  {journal}
  {\bibinfo  {journal} {Phys. Rev. Research}\ }\textbf {\bibinfo {volume}
  {5}},\ \bibinfo {pages} {013149} (\bibinfo {year} {2023})}\BibitemShut
  {NoStop}%
\bibitem [{\citenamefont {Xu}\ \emph {et~al.}(2026)\citenamefont {Xu},
  \citenamefont {Yu}, \citenamefont {Huang},\ and\ \citenamefont
  {Feng}}]{XuPRR2026}%
  \BibitemOpen
  \bibfield  {author} {\bibinfo {author} {\bibfnamefont {A.}~\bibnamefont
  {Xu}}, \bibinfo {author} {\bibfnamefont {N.}~\bibnamefont {Yu}}, \bibinfo
  {author} {\bibfnamefont {D.}~\bibnamefont {Huang}},\ and\ \bibinfo {author}
  {\bibfnamefont {Y.}~\bibnamefont {Feng}},\ }\bibfield  {title} {\bibinfo
  {title} {Identifying liquidlike and gaslike states of dusty plasmas using
  isomorph theory},\ }\href {https://doi.org/10.1103/nygm-3tsj} {\bibfield
  {journal} {\bibinfo  {journal} {Phys. Rev. Res.}\ }\textbf {\bibinfo {volume}
  {8}},\ \bibinfo {pages} {023016} (\bibinfo {year} {2026})}\BibitemShut
  {NoStop}%
\bibitem [{\citenamefont {Mountain}\ and\ \citenamefont
  {Zwanzig}(1966)}]{MountainJCP1966}%
  \BibitemOpen
  \bibfield  {author} {\bibinfo {author} {\bibfnamefont {R.~D.}\ \bibnamefont
  {Mountain}}\ and\ \bibinfo {author} {\bibfnamefont {R.}~\bibnamefont
  {Zwanzig}},\ }\bibfield  {title} {\bibinfo {title} {Shear relaxation times of
  simple fluids},\ }\href {https://doi.org/10.1063/1.1727124} {\bibfield
  {journal} {\bibinfo  {journal} {J. Chem. Phys.}\ }\textbf {\bibinfo {volume}
  {44}},\ \bibinfo {pages} {2777–2779} (\bibinfo {year} {1966})}\BibitemShut
  {NoStop}%
\bibitem [{\citenamefont {Khrapak}\ and\ \citenamefont
  {Khrapak}(2024)}]{KhrapakPRE11_2024}%
  \BibitemOpen
  \bibfield  {author} {\bibinfo {author} {\bibfnamefont {S.~A.}\ \bibnamefont
  {Khrapak}}\ and\ \bibinfo {author} {\bibfnamefont {A.~G.}\ \bibnamefont
  {Khrapak}},\ }\bibfield  {title} {\bibinfo {title} {Quasiuniversal behavior
  of shear relaxation times in simple fluids},\ }\href
  {https://doi.org/10.1103/physreve.110.054101} {\bibfield  {journal} {\bibinfo
   {journal} {Phys. Rev. E}\ }\textbf {\bibinfo {volume} {110}},\ \bibinfo
  {pages} {054101} (\bibinfo {year} {2024})}\BibitemShut {NoStop}%
\bibitem [{\citenamefont {Khrapak}(2021)}]{KhrapakMolecules12_2021}%
  \BibitemOpen
  \bibfield  {author} {\bibinfo {author} {\bibfnamefont {S.~A.}\ \bibnamefont
  {Khrapak}},\ }\bibfield  {title} {\bibinfo {title} {Self-diffusion in simple
  liquids as a random walk process},\ }\href
  {https://doi.org/10.3390/molecules26247499} {\bibfield  {journal} {\bibinfo
  {journal} {Molecules}\ }\textbf {\bibinfo {volume} {26}},\ \bibinfo {pages}
  {7499} (\bibinfo {year} {2021})}\BibitemShut {NoStop}%
\bibitem [{\citenamefont {Khrapak}(2024{\natexlab{a}})}]{KhrapakPhysRep2024}%
  \BibitemOpen
  \bibfield  {author} {\bibinfo {author} {\bibfnamefont {S.}~\bibnamefont
  {Khrapak}},\ }\bibfield  {title} {\bibinfo {title} {Elementary vibrational
  model for transport properties of dense fluids},\ }\href
  {https://doi.org/10.1016/j.physrep.2023.11.004} {\bibfield  {journal}
  {\bibinfo  {journal} {Phys. Rep.}\ }\textbf {\bibinfo {volume} {1050}},\
  \bibinfo {pages} {1} (\bibinfo {year} {2024}{\natexlab{a}})}\BibitemShut
  {NoStop}%
\bibitem [{\citenamefont {Zwanzig}(1983)}]{ZwanzigJCP1983}%
  \BibitemOpen
  \bibfield  {author} {\bibinfo {author} {\bibfnamefont {R.}~\bibnamefont
  {Zwanzig}},\ }\bibfield  {title} {\bibinfo {title} {On the relation between
  self-diffusion and viscosity of liquids},\ }\href
  {https://doi.org/10.1063/1.446338} {\bibfield  {journal} {\bibinfo  {journal}
  {J. Chem. Phys.}\ }\textbf {\bibinfo {volume} {79}},\ \bibinfo {pages} {4507}
  (\bibinfo {year} {1983})}\BibitemShut {NoStop}%
\bibitem [{\citenamefont {Khrapak}(2019{\natexlab{a}})}]{KhrapakMolPhys2019}%
  \BibitemOpen
  \bibfield  {author} {\bibinfo {author} {\bibfnamefont {S.}~\bibnamefont
  {Khrapak}},\ }\bibfield  {title} {\bibinfo {title}
  {Stokes{\textendash}{E}instein relation in simple fluids revisited},\ }\href
  {https://doi.org/10.1080/00268976.2019.1643045} {\bibfield  {journal}
  {\bibinfo  {journal} {Mol. Phys.}\ }\textbf {\bibinfo {volume} {118}},\
  \bibinfo {pages} {e1643045} (\bibinfo {year}
  {2019}{\natexlab{a}})}\BibitemShut {NoStop}%
\bibitem [{\citenamefont {Khrapak}\ and\ \citenamefont
  {Khrapak}(2021)}]{KhrapakPRE10_2021}%
  \BibitemOpen
  \bibfield  {author} {\bibinfo {author} {\bibfnamefont {S.~A.}\ \bibnamefont
  {Khrapak}}\ and\ \bibinfo {author} {\bibfnamefont {A.~G.}\ \bibnamefont
  {Khrapak}},\ }\bibfield  {title} {\bibinfo {title} {Excess entropy and
  {S}tokes-{E}instein relation in simple fluids},\ }\href
  {https://doi.org/10.1103/physreve.104.044110} {\bibfield  {journal} {\bibinfo
   {journal} {Phys. Rev. E}\ }\textbf {\bibinfo {volume} {104}},\ \bibinfo
  {pages} {044110} (\bibinfo {year} {2021})}\BibitemShut {NoStop}%
\bibitem [{\citenamefont {Dyre}(2006)}]{DyreRMP2006}%
  \BibitemOpen
  \bibfield  {author} {\bibinfo {author} {\bibfnamefont {J.~C.}\ \bibnamefont
  {Dyre}},\ }\bibfield  {title} {\bibinfo {title} {Colloquium: {T}he glass
  transition and elastic models of glass-forming liquids},\ }\href
  {https://doi.org/10.1103/revmodphys.78.953} {\bibfield  {journal} {\bibinfo
  {journal} {Rev. Mod. Phys.}\ }\textbf {\bibinfo {volume} {78}},\ \bibinfo
  {pages} {953} (\bibinfo {year} {2006})}\BibitemShut {NoStop}%
\bibitem [{\citenamefont {Dyre}\ and\ \citenamefont
  {Wang}(2012)}]{DyreJCP2012}%
  \BibitemOpen
  \bibfield  {author} {\bibinfo {author} {\bibfnamefont {J.~C.}\ \bibnamefont
  {Dyre}}\ and\ \bibinfo {author} {\bibfnamefont {W.~H.}\ \bibnamefont
  {Wang}},\ }\bibfield  {title} {\bibinfo {title} {The instantaneous shear
  modulus in the shoving model},\ }\href {https://doi.org/10.1063/1.4724102}
  {\bibfield  {journal} {\bibinfo  {journal} {J. Chem. Phys.}\ }\textbf
  {\bibinfo {volume} {136}},\ \bibinfo {pages} {224108} (\bibinfo {year}
  {2012})}\BibitemShut {NoStop}%
\bibitem [{\citenamefont {Khrapak}(2024{\natexlab{b}})}]{KhrapakJCP2024_glass}%
  \BibitemOpen
  \bibfield  {author} {\bibinfo {author} {\bibfnamefont {S.~A.}\ \bibnamefont
  {Khrapak}},\ }\bibfield  {title} {\bibinfo {title} {Shoving model and the
  glass transition in one-component plasma},\ }\href
  {https://doi.org/10.1063/5.0207393} {\bibfield  {journal} {\bibinfo
  {journal} {J. Chem. Phys.}\ }\textbf {\bibinfo {volume} {160}},\ \bibinfo
  {pages} {166101} (\bibinfo {year} {2024}{\natexlab{b}})}\BibitemShut
  {NoStop}%
\bibitem [{\citenamefont {Hosokawa}\ \emph {et~al.}(2009)\citenamefont
  {Hosokawa}, \citenamefont {Inui}, \citenamefont {Kajihara}, \citenamefont
  {Matsuda}, \citenamefont {Ichitsubo}, \citenamefont {Pilgrim}, \citenamefont
  {Sinn}, \citenamefont {Gonz{\'{a}}lez}, \citenamefont {Gonz{\'{a}}lez},
  \citenamefont {Tsutsui},\ and\ \citenamefont {Baron}}]{HosokawaPRL2009}%
  \BibitemOpen
  \bibfield  {author} {\bibinfo {author} {\bibfnamefont {S.}~\bibnamefont
  {Hosokawa}}, \bibinfo {author} {\bibfnamefont {M.}~\bibnamefont {Inui}},
  \bibinfo {author} {\bibfnamefont {Y.}~\bibnamefont {Kajihara}}, \bibinfo
  {author} {\bibfnamefont {K.}~\bibnamefont {Matsuda}}, \bibinfo {author}
  {\bibfnamefont {T.}~\bibnamefont {Ichitsubo}}, \bibinfo {author}
  {\bibfnamefont {W.-C.}\ \bibnamefont {Pilgrim}}, \bibinfo {author}
  {\bibfnamefont {H.}~\bibnamefont {Sinn}}, \bibinfo {author} {\bibfnamefont
  {L.~E.}\ \bibnamefont {Gonz{\'{a}}lez}}, \bibinfo {author} {\bibfnamefont
  {D.~J.}\ \bibnamefont {Gonz{\'{a}}lez}}, \bibinfo {author} {\bibfnamefont
  {S.}~\bibnamefont {Tsutsui}},\ and\ \bibinfo {author} {\bibfnamefont
  {A.~Q.~R.}\ \bibnamefont {Baron}},\ }\bibfield  {title} {\bibinfo {title}
  {Transverse acoustic excitations in liquid {G}a},\ }\href
  {https://doi.org/10.1103/physrevlett.102.105502} {\bibfield  {journal}
  {\bibinfo  {journal} {Phys. Rev. Lett.}\ }\textbf {\bibinfo {volume} {102}},\
  \bibinfo {pages} {105502} (\bibinfo {year} {2009})}\BibitemShut {NoStop}%
\bibitem [{\citenamefont {Hosokawa}\ \emph {et~al.}(2013)\citenamefont
  {Hosokawa}, \citenamefont {Munejiri}, \citenamefont {Inui}, \citenamefont
  {Kajihara}, \citenamefont {Pilgrim}, \citenamefont {Ohmasa}, \citenamefont
  {Tsutsui}, \citenamefont {Baron}, \citenamefont {Shimojo},\ and\
  \citenamefont {Hoshino}}]{HosokawaJPCM2013}%
  \BibitemOpen
  \bibfield  {author} {\bibinfo {author} {\bibfnamefont {S.}~\bibnamefont
  {Hosokawa}}, \bibinfo {author} {\bibfnamefont {S.}~\bibnamefont {Munejiri}},
  \bibinfo {author} {\bibfnamefont {M.}~\bibnamefont {Inui}}, \bibinfo {author}
  {\bibfnamefont {Y.}~\bibnamefont {Kajihara}}, \bibinfo {author}
  {\bibfnamefont {W.-C.}\ \bibnamefont {Pilgrim}}, \bibinfo {author}
  {\bibfnamefont {Y.}~\bibnamefont {Ohmasa}}, \bibinfo {author} {\bibfnamefont
  {S.}~\bibnamefont {Tsutsui}}, \bibinfo {author} {\bibfnamefont {A.~Q.~R.}\
  \bibnamefont {Baron}}, \bibinfo {author} {\bibfnamefont {F.}~\bibnamefont
  {Shimojo}},\ and\ \bibinfo {author} {\bibfnamefont {K.}~\bibnamefont
  {Hoshino}},\ }\bibfield  {title} {\bibinfo {title} {Transverse excitations in
  liquid {S}n},\ }\href {https://doi.org/10.1088/0953-8984/25/11/112101}
  {\bibfield  {journal} {\bibinfo  {journal} {J. Phys.: Condens. Matter}\
  }\textbf {\bibinfo {volume} {25}},\ \bibinfo {pages} {112101} (\bibinfo
  {year} {2013})}\BibitemShut {NoStop}%
\bibitem [{\citenamefont {Hosokawa}\ \emph {et~al.}(2015)\citenamefont
  {Hosokawa}, \citenamefont {Inui}, \citenamefont {Kajihara}, \citenamefont
  {Tsutsui},\ and\ \citenamefont {Baron}}]{HosokawaJPCM2015}%
  \BibitemOpen
  \bibfield  {author} {\bibinfo {author} {\bibfnamefont {S.}~\bibnamefont
  {Hosokawa}}, \bibinfo {author} {\bibfnamefont {M.}~\bibnamefont {Inui}},
  \bibinfo {author} {\bibfnamefont {Y.}~\bibnamefont {Kajihara}}, \bibinfo
  {author} {\bibfnamefont {S.}~\bibnamefont {Tsutsui}},\ and\ \bibinfo {author}
  {\bibfnamefont {A.~Q.~R.}\ \bibnamefont {Baron}},\ }\bibfield  {title}
  {\bibinfo {title} {Transverse excitations in liquid {Fe}, {Cu} and {Zn}},\
  }\href {https://doi.org/10.1088/0953-8984/27/19/194104} {\bibfield  {journal}
  {\bibinfo  {journal} {J. Phys.: Condens. Matter}\ }\textbf {\bibinfo {volume}
  {27}},\ \bibinfo {pages} {194104} (\bibinfo {year} {2015})}\BibitemShut
  {NoStop}%
\bibitem [{\citenamefont {Pramanik}\ \emph {et~al.}(2002)\citenamefont
  {Pramanik}, \citenamefont {Prasad}, \citenamefont {Sen},\ and\ \citenamefont
  {Kaw}}]{PramanikPRL2002}%
  \BibitemOpen
  \bibfield  {author} {\bibinfo {author} {\bibfnamefont {J.}~\bibnamefont
  {Pramanik}}, \bibinfo {author} {\bibfnamefont {G.}~\bibnamefont {Prasad}},
  \bibinfo {author} {\bibfnamefont {A.}~\bibnamefont {Sen}},\ and\ \bibinfo
  {author} {\bibfnamefont {P.~K.}\ \bibnamefont {Kaw}},\ }\bibfield  {title}
  {\bibinfo {title} {Experimental observations of transverse shear waves in
  strongly coupled dusty plasmas},\ }\href
  {https://doi.org/10.1103/physrevlett.88.175001} {\bibfield  {journal}
  {\bibinfo  {journal} {Phys. Rev. Lett.}\ }\textbf {\bibinfo {volume} {88}},\
  \bibinfo {pages} {175001} (\bibinfo {year} {2002})}\BibitemShut {NoStop}%
\bibitem [{\citenamefont {Nunomura}\ \emph {et~al.}(2005)\citenamefont
  {Nunomura}, \citenamefont {Zhdanov}, \citenamefont {Samsonov},\ and\
  \citenamefont {Morfill}}]{NunomuraPRL2005}%
  \BibitemOpen
  \bibfield  {author} {\bibinfo {author} {\bibfnamefont {S.}~\bibnamefont
  {Nunomura}}, \bibinfo {author} {\bibfnamefont {S.}~\bibnamefont {Zhdanov}},
  \bibinfo {author} {\bibfnamefont {D.}~\bibnamefont {Samsonov}},\ and\
  \bibinfo {author} {\bibfnamefont {G.}~\bibnamefont {Morfill}},\ }\bibfield
  {title} {\bibinfo {title} {Wave spectra in solid and liquid complex (dusty)
  plasmas},\ }\href {https://doi.org/10.1103/physrevlett.94.045001} {\bibfield
  {journal} {\bibinfo  {journal} {Phys. Rev. Lett.}\ }\textbf {\bibinfo
  {volume} {94}},\ \bibinfo {pages} {045001} (\bibinfo {year}
  {2005})}\BibitemShut {NoStop}%
\bibitem [{\citenamefont {Bandyopadhyay}\ \emph {et~al.}(2008)\citenamefont
  {Bandyopadhyay}, \citenamefont {Prasad}, \citenamefont {Sen},\ and\
  \citenamefont {Kaw}}]{BandyopadhuayPLA2008}%
  \BibitemOpen
  \bibfield  {author} {\bibinfo {author} {\bibfnamefont {P.}~\bibnamefont
  {Bandyopadhyay}}, \bibinfo {author} {\bibfnamefont {G.}~\bibnamefont
  {Prasad}}, \bibinfo {author} {\bibfnamefont {A.}~\bibnamefont {Sen}},\ and\
  \bibinfo {author} {\bibfnamefont {P.}~\bibnamefont {Kaw}},\ }\bibfield
  {title} {\bibinfo {title} {Driven transverse shear waves in a strongly
  coupled dusty plasma},\ }\href
  {https://doi.org/10.1016/j.physleta.2008.06.051} {\bibfield  {journal}
  {\bibinfo  {journal} {Phys. Lett. A}\ }\textbf {\bibinfo {volume} {372}},\
  \bibinfo {pages} {5467} (\bibinfo {year} {2008})}\BibitemShut {NoStop}%
\bibitem [{\citenamefont {Ohta}\ and\ \citenamefont
  {Hamaguchi}(2000)}]{OhtaPRL2000}%
  \BibitemOpen
  \bibfield  {author} {\bibinfo {author} {\bibfnamefont {H.}~\bibnamefont
  {Ohta}}\ and\ \bibinfo {author} {\bibfnamefont {S.}~\bibnamefont
  {Hamaguchi}},\ }\bibfield  {title} {\bibinfo {title} {Wave dispersion
  relations in {Y}ukawa fluids},\ }\href
  {https://doi.org/10.1103/physrevlett.84.6026} {\bibfield  {journal} {\bibinfo
   {journal} {Phys. Rev. Lett.}\ }\textbf {\bibinfo {volume} {84}},\ \bibinfo
  {pages} {6026} (\bibinfo {year} {2000})}\BibitemShut {NoStop}%
\bibitem [{\citenamefont {Goree}\ \emph {et~al.}(2012)\citenamefont {Goree},
  \citenamefont {Donk{\'{o}}},\ and\ \citenamefont {Hartmann}}]{GoreePRE2012}%
  \BibitemOpen
  \bibfield  {author} {\bibinfo {author} {\bibfnamefont {J.}~\bibnamefont
  {Goree}}, \bibinfo {author} {\bibfnamefont {Z.}~\bibnamefont {Donk{\'{o}}}},\
  and\ \bibinfo {author} {\bibfnamefont {P.}~\bibnamefont {Hartmann}},\
  }\bibfield  {title} {\bibinfo {title} {Cutoff wave number for shear waves and
  {M}axwell relaxation time in {Y}ukawa liquids},\ }\href
  {https://doi.org/10.1103/physreve.85.066401} {\bibfield  {journal} {\bibinfo
  {journal} {Phys. Rev. E}\ }\textbf {\bibinfo {volume} {85}},\ \bibinfo
  {pages} {066401} (\bibinfo {year} {2012})}\BibitemShut {NoStop}%
\bibitem [{\citenamefont {Yang}\ \emph {et~al.}(2017)\citenamefont {Yang},
  \citenamefont {Dove}, \citenamefont {Brazhkin},\ and\ \citenamefont
  {Trachenko}}]{YangPRL2017}%
  \BibitemOpen
  \bibfield  {author} {\bibinfo {author} {\bibfnamefont {C.}~\bibnamefont
  {Yang}}, \bibinfo {author} {\bibfnamefont {M.~T.}\ \bibnamefont {Dove}},
  \bibinfo {author} {\bibfnamefont {V.~V.}\ \bibnamefont {Brazhkin}},\ and\
  \bibinfo {author} {\bibfnamefont {K.}~\bibnamefont {Trachenko}},\ }\bibfield
  {title} {\bibinfo {title} {Emergence and evolution of the k-gap in spectra of
  liquid and supercritical states},\ }\href
  {https://doi.org/10.1103/physrevlett.118.215502} {\bibfield  {journal}
  {\bibinfo  {journal} {Phys. Rev. Lett.}\ }\textbf {\bibinfo {volume} {118}},\
  \bibinfo {pages} {215502} (\bibinfo {year} {2017})}\BibitemShut {NoStop}%
\bibitem [{\citenamefont {Bryk}\ \emph {et~al.}(2017)\citenamefont {Bryk},
  \citenamefont {Huerta}, \citenamefont {Hordiichuk},\ and\ \citenamefont
  {Trokhymchuk}}]{BrykJCP2017}%
  \BibitemOpen
  \bibfield  {author} {\bibinfo {author} {\bibfnamefont {T.}~\bibnamefont
  {Bryk}}, \bibinfo {author} {\bibfnamefont {A.}~\bibnamefont {Huerta}},
  \bibinfo {author} {\bibfnamefont {V.}~\bibnamefont {Hordiichuk}},\ and\
  \bibinfo {author} {\bibfnamefont {A.~D.}\ \bibnamefont {Trokhymchuk}},\
  }\bibfield  {title} {\bibinfo {title} {Non-hydrodynamic transverse collective
  excitations in hard-sphere fluids},\ }\href
  {https://doi.org/10.1063/1.4997640} {\bibfield  {journal} {\bibinfo
  {journal} {J. Chem. Phys.}\ }\textbf {\bibinfo {volume} {147}},\ \bibinfo
  {pages} {064509} (\bibinfo {year} {2017})}\BibitemShut {NoStop}%
\bibitem [{\citenamefont {Kryuchkov}\ \emph {et~al.}(2019)\citenamefont
  {Kryuchkov}, \citenamefont {Mistryukova}, \citenamefont {Brazhkin},\ and\
  \citenamefont {Yurchenko}}]{KryuchkovSciRep2019}%
  \BibitemOpen
  \bibfield  {author} {\bibinfo {author} {\bibfnamefont {N.~P.}\ \bibnamefont
  {Kryuchkov}}, \bibinfo {author} {\bibfnamefont {L.~A.}\ \bibnamefont
  {Mistryukova}}, \bibinfo {author} {\bibfnamefont {V.~V.}\ \bibnamefont
  {Brazhkin}},\ and\ \bibinfo {author} {\bibfnamefont {S.~O.}\ \bibnamefont
  {Yurchenko}},\ }\bibfield  {title} {\bibinfo {title} {Excitation spectra in
  fluids: How to analyze them properly},\ }\href
  {https://doi.org/10.1038/s41598-019-46979-y} {\bibfield  {journal} {\bibinfo
  {journal} {Sci. Rep.}\ }\textbf {\bibinfo {volume} {9}},\ \bibinfo {pages}
  {10483} (\bibinfo {year} {2019})}\BibitemShut {NoStop}%
\bibitem [{\citenamefont {Khrapak}\ \emph {et~al.}(2019)\citenamefont
  {Khrapak}, \citenamefont {Khrapak}, \citenamefont {Kryuchkov},\ and\
  \citenamefont {Yurchenko}}]{KhrapakJCP2019}%
  \BibitemOpen
  \bibfield  {author} {\bibinfo {author} {\bibfnamefont {S.~A.}\ \bibnamefont
  {Khrapak}}, \bibinfo {author} {\bibfnamefont {A.~G.}\ \bibnamefont
  {Khrapak}}, \bibinfo {author} {\bibfnamefont {N.~P.}\ \bibnamefont
  {Kryuchkov}},\ and\ \bibinfo {author} {\bibfnamefont {S.~O.}\ \bibnamefont
  {Yurchenko}},\ }\bibfield  {title} {\bibinfo {title} {Onset of transverse
  (shear) waves in strongly-coupled {Y}ukawa fluids},\ }\href
  {https://doi.org/10.1063/1.5088141} {\bibfield  {journal} {\bibinfo
  {journal} {J. Chem. Phys.}\ }\textbf {\bibinfo {volume} {150}},\ \bibinfo
  {pages} {104503} (\bibinfo {year} {2019})}\BibitemShut {NoStop}%
\bibitem [{\citenamefont {Kryuchkov}\ and\ \citenamefont
  {Yurchenko}(2021)}]{KryuchkovJCP2021}%
  \BibitemOpen
  \bibfield  {author} {\bibinfo {author} {\bibfnamefont {N.~P.}\ \bibnamefont
  {Kryuchkov}}\ and\ \bibinfo {author} {\bibfnamefont {S.~O.}\ \bibnamefont
  {Yurchenko}},\ }\bibfield  {title} {\bibinfo {title} {Collective excitations
  in active fluids: Microflows and breakdown in spectral equipartition of
  kinetic energy},\ }\href {https://doi.org/10.1063/5.0054854} {\bibfield
  {journal} {\bibinfo  {journal} {J. Chem. Phys.}\ }\textbf {\bibinfo {volume}
  {155}},\ \bibinfo {pages} {024902} (\bibinfo {year} {2021})}\BibitemShut
  {NoStop}%
\bibitem [{\citenamefont {Hoover}\ \emph {et~al.}(1970)\citenamefont {Hoover},
  \citenamefont {Ross}, \citenamefont {Johnson}, \citenamefont {Henderson},
  \citenamefont {Barker},\ and\ \citenamefont {Brown}}]{HooverJCP1970}%
  \BibitemOpen
  \bibfield  {author} {\bibinfo {author} {\bibfnamefont {W.~G.}\ \bibnamefont
  {Hoover}}, \bibinfo {author} {\bibfnamefont {M.}~\bibnamefont {Ross}},
  \bibinfo {author} {\bibfnamefont {K.~W.}\ \bibnamefont {Johnson}}, \bibinfo
  {author} {\bibfnamefont {D.}~\bibnamefont {Henderson}}, \bibinfo {author}
  {\bibfnamefont {J.~A.}\ \bibnamefont {Barker}},\ and\ \bibinfo {author}
  {\bibfnamefont {B.~C.}\ \bibnamefont {Brown}},\ }\bibfield  {title} {\bibinfo
  {title} {Soft-sphere equation of state},\ }\href
  {https://doi.org/10.1063/1.1672728} {\bibfield  {journal} {\bibinfo
  {journal} {J. Chem. Phys.}\ }\textbf {\bibinfo {volume} {52}},\ \bibinfo
  {pages} {4931–4941} (\bibinfo {year} {1970})}\BibitemShut {NoStop}%
\bibitem [{\citenamefont {Hoover}\ \emph {et~al.}(1971)\citenamefont {Hoover},
  \citenamefont {Gray},\ and\ \citenamefont {Johnson}}]{HooverJCP1971}%
  \BibitemOpen
  \bibfield  {author} {\bibinfo {author} {\bibfnamefont {W.~G.}\ \bibnamefont
  {Hoover}}, \bibinfo {author} {\bibfnamefont {S.~G.}\ \bibnamefont {Gray}},\
  and\ \bibinfo {author} {\bibfnamefont {K.~W.}\ \bibnamefont {Johnson}},\
  }\bibfield  {title} {\bibinfo {title} {Thermodynamic properties of the fluid
  and solid phases for inverse power potentials},\ }\href
  {https://doi.org/10.1063/1.1676196} {\bibfield  {journal} {\bibinfo
  {journal} {J. Chem. Phys.}\ }\textbf {\bibinfo {volume} {55}},\ \bibinfo
  {pages} {1128–1136} (\bibinfo {year} {1971})}\BibitemShut {NoStop}%
\bibitem [{\citenamefont {{Tsytovich}}(1997)}]{TsytovichUFN1997}%
  \BibitemOpen
  \bibfield  {author} {\bibinfo {author} {\bibfnamefont {V.}~\bibnamefont
  {{Tsytovich}}},\ }\bibfield  {title} {\bibinfo {title} {{Dust plasma
  crystals, drops, and clouds.}},\ }\href
  {https://doi.org/10.1070/PU1997v040n01ABEH000201} {\bibfield  {journal}
  {\bibinfo  {journal} {Phys.-Usp.}\ }\textbf {\bibinfo {volume} {40}},\
  \bibinfo {pages} {53} (\bibinfo {year} {1997})}\BibitemShut {NoStop}%
\bibitem [{\citenamefont {Fortov}\ \emph {et~al.}(2004)\citenamefont {Fortov},
  \citenamefont {Khrapak}, \citenamefont {Khrapak}, \citenamefont {Molotkov},\
  and\ \citenamefont {Petrov}}]{FortovUFN}%
  \BibitemOpen
  \bibfield  {author} {\bibinfo {author} {\bibfnamefont {V.~E.}\ \bibnamefont
  {Fortov}}, \bibinfo {author} {\bibfnamefont {A.~G.}\ \bibnamefont {Khrapak}},
  \bibinfo {author} {\bibfnamefont {S.~A.}\ \bibnamefont {Khrapak}}, \bibinfo
  {author} {\bibfnamefont {V.~I.}\ \bibnamefont {Molotkov}},\ and\ \bibinfo
  {author} {\bibfnamefont {O.~F.}\ \bibnamefont {Petrov}},\ }\bibfield  {title}
  {\bibinfo {title} {Dusty plasmas},\ }\href
  {https://doi.org/10.3367/ufnr.0174.200405b.0495} {\bibfield  {journal}
  {\bibinfo  {journal} {Phys.-Usp.}\ }\textbf {\bibinfo {volume} {47}},\
  \bibinfo {pages} {447 } (\bibinfo {year} {2004})}\BibitemShut {NoStop}%
\bibitem [{\citenamefont {Fortov}\ \emph {et~al.}(2005)\citenamefont {Fortov},
  \citenamefont {Ivlev}, \citenamefont {Khrapak}, \citenamefont {Khrapak},\
  and\ \citenamefont {Morfill}}]{FortovPR}%
  \BibitemOpen
  \bibfield  {author} {\bibinfo {author} {\bibfnamefont {V.~E.}\ \bibnamefont
  {Fortov}}, \bibinfo {author} {\bibfnamefont {A.~V.}\ \bibnamefont {Ivlev}},
  \bibinfo {author} {\bibfnamefont {S.~A.}\ \bibnamefont {Khrapak}}, \bibinfo
  {author} {\bibfnamefont {A.~G.}\ \bibnamefont {Khrapak}},\ and\ \bibinfo
  {author} {\bibfnamefont {G.~E.}\ \bibnamefont {Morfill}},\ }\bibfield
  {title} {\bibinfo {title} {Complex (dusty) plasmas: Current status, open
  issues, perspectives},\ }\href@noop {} {\bibfield  {journal} {\bibinfo
  {journal} {Phys. Rep.}\ }\textbf {\bibinfo {volume} {421}},\ \bibinfo {pages}
  {1} (\bibinfo {year} {2005})}\BibitemShut {NoStop}%
\bibitem [{\citenamefont {Shukla}\ and\ \citenamefont
  {Eliasson}(2009)}]{ShuklaRMP2009}%
  \BibitemOpen
  \bibfield  {author} {\bibinfo {author} {\bibfnamefont {P.~K.}\ \bibnamefont
  {Shukla}}\ and\ \bibinfo {author} {\bibfnamefont {B.}~\bibnamefont
  {Eliasson}},\ }\bibfield  {title} {\bibinfo {title} {Colloquium: Fundamentals
  of dust-plasma interactions},\ }\href
  {https://doi.org/10.1103/revmodphys.81.25} {\bibfield  {journal} {\bibinfo
  {journal} {Rev. Mod. Phys.}\ }\textbf {\bibinfo {volume} {81}},\ \bibinfo
  {pages} {25–44} (\bibinfo {year} {2009})}\BibitemShut {NoStop}%
\bibitem [{\citenamefont {Morfill}\ and\ \citenamefont
  {Ivlev}(2009)}]{MorfillRMP2009}%
  \BibitemOpen
  \bibfield  {author} {\bibinfo {author} {\bibfnamefont {G.~E.}\ \bibnamefont
  {Morfill}}\ and\ \bibinfo {author} {\bibfnamefont {A.~V.}\ \bibnamefont
  {Ivlev}},\ }\bibfield  {title} {\bibinfo {title} {Complex plasmas: An
  interdisciplinary research field},\ }\href
  {https://doi.org/10.1103/revmodphys.81.1353} {\bibfield  {journal} {\bibinfo
  {journal} {Rev. Mod. Phys.}\ }\textbf {\bibinfo {volume} {81}},\ \bibinfo
  {pages} {1353–1404} (\bibinfo {year} {2009})}\BibitemShut {NoStop}%
\bibitem [{\citenamefont {Fortov}\ and\ \citenamefont
  {Morfill}(2010)}]{FortovBook}%
  \BibitemOpen
  \bibfield  {author} {\bibinfo {author} {\bibfnamefont {V.~E.}\ \bibnamefont
  {Fortov}}\ and\ \bibinfo {author} {\bibfnamefont {G.~E.}\ \bibnamefont
  {Morfill}},\ }\href@noop {} {\emph {\bibinfo {title} {Complex and Dusty
  Plasmas - From Laboratory to Space}}}\ (\bibinfo  {publisher} {CRC Press
  LLC},\ \bibinfo {address} {Boca Raton},\ \bibinfo {year} {2010})\BibitemShut
  {NoStop}%
\bibitem [{\citenamefont {Ivlev}\ \emph {et~al.}(2012)\citenamefont {Ivlev},
  \citenamefont {L\"owen}, \citenamefont {Morfill},\ and\ \citenamefont
  {Royall}}]{IvlevBook}%
  \BibitemOpen
  \bibfield  {author} {\bibinfo {author} {\bibfnamefont {A.}~\bibnamefont
  {Ivlev}}, \bibinfo {author} {\bibfnamefont {H.}~\bibnamefont {L\"owen}},
  \bibinfo {author} {\bibfnamefont {G.}~\bibnamefont {Morfill}},\ and\ \bibinfo
  {author} {\bibfnamefont {C.~P.}\ \bibnamefont {Royall}},\ }\href@noop {}
  {\emph {\bibinfo {title} {Complex Plasmas and Colloidal Dispersions:
  Particle-Resolved Studies of Classical Liquids and Solids}}}\ (\bibinfo
  {publisher} {World Scientific},\ \bibinfo {year} {2012})\BibitemShut
  {NoStop}%
\bibitem [{\citenamefont {Lennard-Jones}(1924)}]{LennardJones1924}%
  \BibitemOpen
  \bibfield  {author} {\bibinfo {author} {\bibfnamefont {J.~E.}\ \bibnamefont
  {Lennard-Jones}},\ }\bibfield  {title} {\bibinfo {title} {On the
  determination of molecular fields.—{I}. {F}rom the variation of the
  viscosity of a gas with temperature},\ }\href
  {https://doi.org/10.1098/rspa.1924.0081} {\bibfield  {journal} {\bibinfo
  {journal} {Proc. R. Soc. Lond. A}\ }\textbf {\bibinfo {volume} {106}},\
  \bibinfo {pages} {441–462} (\bibinfo {year} {1924})}\BibitemShut {NoStop}%
\bibitem [{\citenamefont {Zwanzig}\ and\ \citenamefont
  {Mountain}(1965)}]{ZwanzigJCP1965}%
  \BibitemOpen
  \bibfield  {author} {\bibinfo {author} {\bibfnamefont {R.}~\bibnamefont
  {Zwanzig}}\ and\ \bibinfo {author} {\bibfnamefont {R.~D.}\ \bibnamefont
  {Mountain}},\ }\bibfield  {title} {\bibinfo {title} {High-frequency elastic
  moduli of simple fluids},\ }\href {https://doi.org/10.1063/1.1696718}
  {\bibfield  {journal} {\bibinfo  {journal} {J. Chem. Phys.}\ }\textbf
  {\bibinfo {volume} {43}},\ \bibinfo {pages} {4464} (\bibinfo {year}
  {1965})}\BibitemShut {NoStop}%
\bibitem [{\citenamefont {Balucani}\ and\ \citenamefont
  {Zoppi}(1994)}]{BalucaniBook}%
  \BibitemOpen
  \bibfield  {author} {\bibinfo {author} {\bibfnamefont {U.}~\bibnamefont
  {Balucani}}\ and\ \bibinfo {author} {\bibfnamefont {M.}~\bibnamefont
  {Zoppi}},\ }\href@noop {} {\emph {\bibinfo {title} {Dynamics of the Liquid
  State}}}\ (\bibinfo  {publisher} {Clarendon Press},\ \bibinfo {address}
  {Oxford},\ \bibinfo {year} {1994})\BibitemShut {NoStop}%
\bibitem [{\citenamefont {Bolmatov}\ \emph {et~al.}(2015)\citenamefont
  {Bolmatov}, \citenamefont {Zhernenkov}, \citenamefont {Zav'yalov},
  \citenamefont {Stoupin}, \citenamefont {Cai},\ and\ \citenamefont
  {Cunsolo}}]{BolmatovPCL2015}%
  \BibitemOpen
  \bibfield  {author} {\bibinfo {author} {\bibfnamefont {D.}~\bibnamefont
  {Bolmatov}}, \bibinfo {author} {\bibfnamefont {M.}~\bibnamefont
  {Zhernenkov}}, \bibinfo {author} {\bibfnamefont {D.}~\bibnamefont
  {Zav'yalov}}, \bibinfo {author} {\bibfnamefont {S.}~\bibnamefont {Stoupin}},
  \bibinfo {author} {\bibfnamefont {Y.~Q.}\ \bibnamefont {Cai}},\ and\ \bibinfo
  {author} {\bibfnamefont {A.}~\bibnamefont {Cunsolo}},\ }\bibfield  {title}
  {\bibinfo {title} {Revealing the mechanism of the viscous-to-elastic
  crossover in liquids},\ }\href {https://doi.org/10.1021/acs.jpclett.5b01338}
  {\bibfield  {journal} {\bibinfo  {journal} {J. Phys. Chem. Lett.}\ }\textbf
  {\bibinfo {volume} {6}},\ \bibinfo {pages} {3048} (\bibinfo {year}
  {2015})}\BibitemShut {NoStop}%
\bibitem [{\citenamefont {Trachenko}(2023)}]{TrachenkoBook}%
  \BibitemOpen
  \bibfield  {author} {\bibinfo {author} {\bibfnamefont {K.}~\bibnamefont
  {Trachenko}},\ }\href@noop {} {\emph {\bibinfo {title} {Theory of liquids:
  {F}rom excitations to Thermodynamics}}}\ (\bibinfo  {publisher} {Cambridge
  University Press},\ \bibinfo {address} {Cambridge, England},\ \bibinfo {year}
  {2023})\BibitemShut {NoStop}%
\bibitem [{\citenamefont {Hansen}\ and\ \citenamefont
  {McDonald}(2006)}]{HansenBook}%
  \BibitemOpen
  \bibfield  {author} {\bibinfo {author} {\bibfnamefont {J.-P.}\ \bibnamefont
  {Hansen}}\ and\ \bibinfo {author} {\bibfnamefont {I.~R.}\ \bibnamefont
  {McDonald}},\ }\href@noop {} {\emph {\bibinfo {title} {Theory of Simple
  Liquids -}}}\ (\bibinfo  {publisher} {Elsevier},\ \bibinfo {address}
  {Amsterdam},\ \bibinfo {year} {2006})\BibitemShut {NoStop}%
\bibitem [{\citenamefont {Heyes}\ and\ \citenamefont
  {Aston}(1994)}]{HeyesJCP1994}%
  \BibitemOpen
  \bibfield  {author} {\bibinfo {author} {\bibfnamefont {D.~M.}\ \bibnamefont
  {Heyes}}\ and\ \bibinfo {author} {\bibfnamefont {P.~J.}\ \bibnamefont
  {Aston}},\ }\bibfield  {title} {\bibinfo {title} {Elastic moduli of simple
  fluids with steeply repulsive potentials},\ }\href
  {https://doi.org/10.1063/1.466511} {\bibfield  {journal} {\bibinfo  {journal}
  {J. Chem. Phys.}\ }\textbf {\bibinfo {volume} {100}},\ \bibinfo {pages}
  {2149–2153} (\bibinfo {year} {1994})}\BibitemShut {NoStop}%
\bibitem [{\citenamefont {Khrapak}\ \emph {et~al.}(2017)\citenamefont
  {Khrapak}, \citenamefont {Klumov},\ and\ \citenamefont
  {Couedel}}]{KhrapakSciRep2017}%
  \BibitemOpen
  \bibfield  {author} {\bibinfo {author} {\bibfnamefont {S.}~\bibnamefont
  {Khrapak}}, \bibinfo {author} {\bibfnamefont {B.}~\bibnamefont {Klumov}},\
  and\ \bibinfo {author} {\bibfnamefont {L.}~\bibnamefont {Couedel}},\
  }\bibfield  {title} {\bibinfo {title} {Collective modes in simple melts:
  Transition from soft spheres to the hard sphere limit},\ }\href
  {https://doi.org/10.1038/s41598-017-08429-5} {\bibfield  {journal} {\bibinfo
  {journal} {Sci. Rep.}\ }\textbf {\bibinfo {volume} {7}},\ \bibinfo {pages}
  {7985} (\bibinfo {year} {2017})}\BibitemShut {NoStop}%
\bibitem [{\citenamefont {Heyes}\ \emph {et~al.}(2025)\citenamefont {Heyes},
  \citenamefont {Dini}, \citenamefont {Pieprzyk},\ and\ \citenamefont
  {Brańka}}]{HeyesJCP2025}%
  \BibitemOpen
  \bibfield  {author} {\bibinfo {author} {\bibfnamefont {D.~M.}\ \bibnamefont
  {Heyes}}, \bibinfo {author} {\bibfnamefont {D.}~\bibnamefont {Dini}},
  \bibinfo {author} {\bibfnamefont {S.}~\bibnamefont {Pieprzyk}},\ and\
  \bibinfo {author} {\bibfnamefont {A.~C.}\ \bibnamefont {Brańka}},\
  }\bibfield  {title} {\bibinfo {title} {Equations of state and excess entropy
  of repulsive inverse power particle potential fluids with variable
  stiffness},\ }\href {https://doi.org/10.1063/5.0288082} {\bibfield  {journal}
  {\bibinfo  {journal} {J. Chem. Phys.}\ }\textbf {\bibinfo {volume} {163}},\
  \bibinfo {pages} {114502} (\bibinfo {year} {2025})}\BibitemShut {NoStop}%
\bibitem [{\citenamefont {Baus}\ and\ \citenamefont
  {Hansen}(1980)}]{BausPR1980}%
  \BibitemOpen
  \bibfield  {author} {\bibinfo {author} {\bibfnamefont {M.}~\bibnamefont
  {Baus}}\ and\ \bibinfo {author} {\bibfnamefont {J.~P.}\ \bibnamefont
  {Hansen}},\ }\bibfield  {title} {\bibinfo {title} {Statistical mechanics of
  simple {C}oulomb systems},\ }\href
  {https://doi.org/10.1016/0370-1573(80)90022-8} {\bibfield  {journal}
  {\bibinfo  {journal} {Phys. Rep.}\ }\textbf {\bibinfo {volume} {59}},\
  \bibinfo {pages} {1} (\bibinfo {year} {1980})}\BibitemShut {NoStop}%
\bibitem [{\citenamefont {Golden}\ and\ \citenamefont
  {Kalman}(2000)}]{GoldenPoP2000}%
  \BibitemOpen
  \bibfield  {author} {\bibinfo {author} {\bibfnamefont {K.~I.}\ \bibnamefont
  {Golden}}\ and\ \bibinfo {author} {\bibfnamefont {G.~J.}\ \bibnamefont
  {Kalman}},\ }\bibfield  {title} {\bibinfo {title} {Quasilocalized charge
  approximation in strongly coupled plasma physics},\ }\href
  {https://doi.org/10.1063/1.873814} {\bibfield  {journal} {\bibinfo  {journal}
  {Phys. Plasmas}\ }\textbf {\bibinfo {volume} {7}},\ \bibinfo {pages} {14}
  (\bibinfo {year} {2000})}\BibitemShut {NoStop}%
\bibitem [{\citenamefont {Khrapak}\ \emph {et~al.}(2016)\citenamefont
  {Khrapak}, \citenamefont {Klumov}, \citenamefont {Couedel},\ and\
  \citenamefont {Thomas}}]{KhrapakPoP2016}%
  \BibitemOpen
  \bibfield  {author} {\bibinfo {author} {\bibfnamefont {S.~A.}\ \bibnamefont
  {Khrapak}}, \bibinfo {author} {\bibfnamefont {B.}~\bibnamefont {Klumov}},
  \bibinfo {author} {\bibfnamefont {L.}~\bibnamefont {Couedel}},\ and\ \bibinfo
  {author} {\bibfnamefont {H.~M.}\ \bibnamefont {Thomas}},\ }\bibfield  {title}
  {\bibinfo {title} {On the long-waves dispersion in {Y}ukawa systems},\ }\href
  {https://doi.org/10.1063/1.4942169} {\bibfield  {journal} {\bibinfo
  {journal} {Phys. Plasmas}\ }\textbf {\bibinfo {volume} {23}},\ \bibinfo
  {pages} {023702} (\bibinfo {year} {2016})}\BibitemShut {NoStop}%
\bibitem [{\citenamefont {Khrapak}\ and\ \citenamefont
  {Khrapak}(2016)}]{KhrapakCPP2016}%
  \BibitemOpen
  \bibfield  {author} {\bibinfo {author} {\bibfnamefont {S.~A.}\ \bibnamefont
  {Khrapak}}\ and\ \bibinfo {author} {\bibfnamefont {A.~G.}\ \bibnamefont
  {Khrapak}},\ }\bibfield  {title} {\bibinfo {title} {Internal energy of the
  classical two- and three-dimensional one-component-plasma},\ }\href
  {https://doi.org/10.1002/ctpp.201500104} {\bibfield  {journal} {\bibinfo
  {journal} {Contrib. Plasma Phys.}\ }\textbf {\bibinfo {volume} {56}},\
  \bibinfo {pages} {270} (\bibinfo {year} {2016})}\BibitemShut {NoStop}%
\bibitem [{\citenamefont {Khrapak}(2020{\natexlab{a}})}]{KhrapakMolecules2020}%
  \BibitemOpen
  \bibfield  {author} {\bibinfo {author} {\bibfnamefont {S.~A.}\ \bibnamefont
  {Khrapak}},\ }\bibfield  {title} {\bibinfo {title} {Sound velocities of
  {L}ennard-{J}ones systems near the liquid-solid phase transition},\ }\href
  {https://doi.org/10.3390/molecules25153498} {\bibfield  {journal} {\bibinfo
  {journal} {Molecules}\ }\textbf {\bibinfo {volume} {25}},\ \bibinfo {pages}
  {3498} (\bibinfo {year} {2020}{\natexlab{a}})}\BibitemShut {NoStop}%
\bibitem [{\citenamefont {Khrapak}\ and\ \citenamefont
  {Khrapak}(2023{\natexlab{a}})}]{KhrapakPRE12_2023}%
  \BibitemOpen
  \bibfield  {author} {\bibinfo {author} {\bibfnamefont {S.~A.}\ \bibnamefont
  {Khrapak}}\ and\ \bibinfo {author} {\bibfnamefont {A.~G.}\ \bibnamefont
  {Khrapak}},\ }\bibfield  {title} {\bibinfo {title} {Vibrational model for
  thermal conductivity of {L}ennard-{J}ones fluids: {A}pplicability domain and
  accuracy level},\ }\href {https://doi.org/10.1103/physreve.108.064129}
  {\bibfield  {journal} {\bibinfo  {journal} {Phy. Rev. E}\ }\textbf {\bibinfo
  {volume} {108}},\ \bibinfo {pages} {064129} (\bibinfo {year}
  {2023}{\natexlab{a}})}\BibitemShut {NoStop}%
\bibitem [{\citenamefont {Thol}\ \emph {et~al.}(2016)\citenamefont {Thol},
  \citenamefont {Rutkai}, \citenamefont {K\"{o}ster}, \citenamefont {Lustig},
  \citenamefont {Span},\ and\ \citenamefont {Vrabec}}]{Thol2016}%
  \BibitemOpen
  \bibfield  {author} {\bibinfo {author} {\bibfnamefont {M.}~\bibnamefont
  {Thol}}, \bibinfo {author} {\bibfnamefont {G.}~\bibnamefont {Rutkai}},
  \bibinfo {author} {\bibfnamefont {A.}~\bibnamefont {K\"{o}ster}}, \bibinfo
  {author} {\bibfnamefont {R.}~\bibnamefont {Lustig}}, \bibinfo {author}
  {\bibfnamefont {R.}~\bibnamefont {Span}},\ and\ \bibinfo {author}
  {\bibfnamefont {J.}~\bibnamefont {Vrabec}},\ }\bibfield  {title} {\bibinfo
  {title} {Equation of state for the {L}ennard-{J}ones fluid},\ }\href
  {https://doi.org/10.1063/1.4945000} {\bibfield  {journal} {\bibinfo
  {journal} {J. Phys. Chem. Ref. Data}\ }\textbf {\bibinfo {volume} {45}},\
  \bibinfo {pages} {023101} (\bibinfo {year} {2016})}\BibitemShut {NoStop}%
\bibitem [{\citenamefont {Frisch}(1966)}]{Frisch1966}%
  \BibitemOpen
  \bibfield  {author} {\bibinfo {author} {\bibfnamefont {H.~L.}\ \bibnamefont
  {Frisch}},\ }\bibfield  {title} {\bibinfo {title} {High frequency linear
  response of classical fluids},\ }\href
  {https://doi.org/10.1103/physicsphysiquefizika.2.209} {\bibfield  {journal}
  {\bibinfo  {journal} {Phys.}\ }\textbf {\bibinfo {volume} {2}},\ \bibinfo
  {pages} {209–215} (\bibinfo {year} {1966})}\BibitemShut {NoStop}%
\bibitem [{\citenamefont {Khrapak}(2019{\natexlab{b}})}]{KhrapakPRE09_2019}%
  \BibitemOpen
  \bibfield  {author} {\bibinfo {author} {\bibfnamefont {S.}~\bibnamefont
  {Khrapak}},\ }\bibfield  {title} {\bibinfo {title} {Elastic properties of
  dense hard-sphere fluids},\ }\href
  {https://doi.org/10.1103/physreve.100.032138} {\bibfield  {journal} {\bibinfo
   {journal} {Phys. Rev. E}\ }\textbf {\bibinfo {volume} {100}},\ \bibinfo
  {pages} {032138} (\bibinfo {year} {2019}{\natexlab{b}})}\BibitemShut
  {NoStop}%
\bibitem [{\citenamefont {Khrapak}\ \emph {et~al.}(2021)\citenamefont
  {Khrapak}, \citenamefont {Kryuchkov}, \citenamefont {Mistryukova},\ and\
  \citenamefont {Yurchenko}}]{KhrapakPRE05_2021}%
  \BibitemOpen
  \bibfield  {author} {\bibinfo {author} {\bibfnamefont {S.}~\bibnamefont
  {Khrapak}}, \bibinfo {author} {\bibfnamefont {N.~P.}\ \bibnamefont
  {Kryuchkov}}, \bibinfo {author} {\bibfnamefont {L.~A.}\ \bibnamefont
  {Mistryukova}},\ and\ \bibinfo {author} {\bibfnamefont {S.~O.}\ \bibnamefont
  {Yurchenko}},\ }\bibfield  {title} {\bibinfo {title} {From soft- to
  hard-sphere fluids: Crossover evidenced by high-frequency elastic moduli},\
  }\href {https://doi.org/10.1103/physreve.103.052117} {\bibfield  {journal}
  {\bibinfo  {journal} {Phys. Rev. E}\ }\textbf {\bibinfo {volume} {103}},\
  \bibinfo {pages} {052117} (\bibinfo {year} {2021})}\BibitemShut {NoStop}%
\bibitem [{\citenamefont {Miller}(1969)}]{MillerJCP1969}%
  \BibitemOpen
  \bibfield  {author} {\bibinfo {author} {\bibfnamefont {B.~N.}\ \bibnamefont
  {Miller}},\ }\bibfield  {title} {\bibinfo {title} {Elastic moduli of a fluid
  of rigid spheres},\ }\href {https://doi.org/10.1063/1.1671437} {\bibfield
  {journal} {\bibinfo  {journal} {J. Chem. Phys.}\ }\textbf {\bibinfo {volume}
  {50}},\ \bibinfo {pages} {2733} (\bibinfo {year} {1969})}\BibitemShut
  {NoStop}%
\bibitem [{\citenamefont {Tao}\ \emph {et~al.}(1992)\citenamefont {Tao},
  \citenamefont {Song},\ and\ \citenamefont {Mason}}]{TaoPRA1992}%
  \BibitemOpen
  \bibfield  {author} {\bibinfo {author} {\bibfnamefont {F.-M.}\ \bibnamefont
  {Tao}}, \bibinfo {author} {\bibfnamefont {Y.}~\bibnamefont {Song}},\ and\
  \bibinfo {author} {\bibfnamefont {E.~A.}\ \bibnamefont {Mason}},\ }\bibfield
  {title} {\bibinfo {title} {Derivative of the hard-sphere radial distribution
  function at contact},\ }\href {https://doi.org/10.1103/physreva.46.8007}
  {\bibfield  {journal} {\bibinfo  {journal} {Phys. Rev. A}\ }\textbf {\bibinfo
  {volume} {46}},\ \bibinfo {pages} {8007} (\bibinfo {year}
  {1992})}\BibitemShut {NoStop}%
\bibitem [{\citenamefont {Carnahan}\ and\ \citenamefont
  {Starling}(1969)}]{CarnahanJCP1969}%
  \BibitemOpen
  \bibfield  {author} {\bibinfo {author} {\bibfnamefont {N.~F.}\ \bibnamefont
  {Carnahan}}\ and\ \bibinfo {author} {\bibfnamefont {K.~E.}\ \bibnamefont
  {Starling}},\ }\bibfield  {title} {\bibinfo {title} {Equation of state for
  nonattracting rigid spheres},\ }\href {https://doi.org/10.1063/1.1672048}
  {\bibfield  {journal} {\bibinfo  {journal} {J. Chem. Phys.}\ }\textbf
  {\bibinfo {volume} {51}},\ \bibinfo {pages} {635} (\bibinfo {year}
  {1969})}\BibitemShut {NoStop}%
\bibitem [{\citenamefont {Khrapak}\ and\ \citenamefont
  {Khrapak}(2026)}]{OCP_Variational}%
  \BibitemOpen
  \bibfield  {author} {\bibinfo {author} {\bibfnamefont {S.~A.}\ \bibnamefont
  {Khrapak}}\ and\ \bibinfo {author} {\bibfnamefont {A.~G.}\ \bibnamefont
  {Khrapak}},\ }\bibfield  {title} {\bibinfo {title} {Excess energy of strongly
  coupled one-component plasma from variational approach},\ }\href@noop {}
  {\bibfield  {journal} {\bibinfo  {journal} {J. Exp. Theor. Phys.}\ }\textbf
  {\bibinfo {volume} {169}},\ \bibinfo {pages} {355} (\bibinfo {year}
  {2026})}\BibitemShut {NoStop}%
\bibitem [{\citenamefont {Khrapak}\ and\ \citenamefont
  {Khrapak}(tted)}]{YukawaSubmitted}%
  \BibitemOpen
  \bibfield  {author} {\bibinfo {author} {\bibfnamefont {S.~A.}\ \bibnamefont
  {Khrapak}}\ and\ \bibinfo {author} {\bibfnamefont {A.~G.}\ \bibnamefont
  {Khrapak}},\ }\bibfield  {title} {\bibinfo {title} {Variational approach to
  thermodynamics and elastic moduli of strongly coupled {Y}ukawa fluids},\
  }\href@noop {} {\bibfield  {journal} {\bibinfo  {journal} {Phys. Rev. E}\ }
  (\bibinfo {year} {2026, submitted})}\BibitemShut {NoStop}%
\bibitem [{\citenamefont {Khrapak}(2024{\natexlab{c}})}]{KhrapakPRE09_2024}%
  \BibitemOpen
  \bibfield  {author} {\bibinfo {author} {\bibfnamefont {S.~A.}\ \bibnamefont
  {Khrapak}},\ }\bibfield  {title} {\bibinfo {title} {Entropy of strongly
  coupled {Y}ukawa fluids},\ }\href
  {https://doi.org/10.1103/physreve.110.034602} {\bibfield  {journal} {\bibinfo
   {journal} {Phys. Rev. E}\ }\textbf {\bibinfo {volume} {110}},\ \bibinfo
  {pages} {034602} (\bibinfo {year} {2024}{\natexlab{c}})}\BibitemShut
  {NoStop}%
\bibitem [{\citenamefont {Hamaguchi}\ \emph {et~al.}(1997)\citenamefont
  {Hamaguchi}, \citenamefont {Farouki},\ and\ \citenamefont
  {Dubin}}]{HamaguchiPRE1997}%
  \BibitemOpen
  \bibfield  {author} {\bibinfo {author} {\bibfnamefont {S.}~\bibnamefont
  {Hamaguchi}}, \bibinfo {author} {\bibfnamefont {R.~T.}\ \bibnamefont
  {Farouki}},\ and\ \bibinfo {author} {\bibfnamefont {D.~H.~E.}\ \bibnamefont
  {Dubin}},\ }\bibfield  {title} {\bibinfo {title} {Triple point of {Y}ukawa
  systems},\ }\href {https://doi.org/10.1103/physreve.56.4671} {\bibfield
  {journal} {\bibinfo  {journal} {Phys. Rev. E}\ }\textbf {\bibinfo {volume}
  {56}},\ \bibinfo {pages} {4671} (\bibinfo {year} {1997})}\BibitemShut
  {NoStop}%
\bibitem [{\citenamefont {Morkel}\ \emph {et~al.}(1993)\citenamefont {Morkel},
  \citenamefont {Bodensteiner},\ and\ \citenamefont
  {Gemperlein}}]{MorkelPRE1993}%
  \BibitemOpen
  \bibfield  {author} {\bibinfo {author} {\bibfnamefont {C.}~\bibnamefont
  {Morkel}}, \bibinfo {author} {\bibfnamefont {T.}~\bibnamefont
  {Bodensteiner}},\ and\ \bibinfo {author} {\bibfnamefont {H.}~\bibnamefont
  {Gemperlein}},\ }\bibfield  {title} {\bibinfo {title} {Zero-sound-like modes
  in simple liquid metals},\ }\href {https://doi.org/10.1103/physreve.47.2575}
  {\bibfield  {journal} {\bibinfo  {journal} {Phys. Rev. E}\ }\textbf {\bibinfo
  {volume} {47}},\ \bibinfo {pages} {2575} (\bibinfo {year}
  {1993})}\BibitemShut {NoStop}%
\bibitem [{\citenamefont {Rahman}\ and\ \citenamefont
  {Stillinger}(1974)}]{RahmanPRA1974}%
  \BibitemOpen
  \bibfield  {author} {\bibinfo {author} {\bibfnamefont {A.}~\bibnamefont
  {Rahman}}\ and\ \bibinfo {author} {\bibfnamefont {F.~H.}\ \bibnamefont
  {Stillinger}},\ }\bibfield  {title} {\bibinfo {title} {Propagation of sound
  in water. {A} molecular-dynamics study},\ }\href
  {https://doi.org/10.1103/physreva.10.368} {\bibfield  {journal} {\bibinfo
  {journal} {Phys. Rev. A}\ }\textbf {\bibinfo {volume} {10}},\ \bibinfo
  {pages} {368–378} (\bibinfo {year} {1974})}\BibitemShut {NoStop}%
\bibitem [{\citenamefont {Pontecorvo}\ \emph {et~al.}(2005)\citenamefont
  {Pontecorvo}, \citenamefont {Krisch}, \citenamefont {Cunsolo}, \citenamefont
  {Monaco}, \citenamefont {Mermet}, \citenamefont {Verbeni}, \citenamefont
  {Sette},\ and\ \citenamefont {Ruocco}}]{PontecorvoPRE2005}%
  \BibitemOpen
  \bibfield  {author} {\bibinfo {author} {\bibfnamefont {E.}~\bibnamefont
  {Pontecorvo}}, \bibinfo {author} {\bibfnamefont {M.}~\bibnamefont {Krisch}},
  \bibinfo {author} {\bibfnamefont {A.}~\bibnamefont {Cunsolo}}, \bibinfo
  {author} {\bibfnamefont {G.}~\bibnamefont {Monaco}}, \bibinfo {author}
  {\bibfnamefont {A.}~\bibnamefont {Mermet}}, \bibinfo {author} {\bibfnamefont
  {R.}~\bibnamefont {Verbeni}}, \bibinfo {author} {\bibfnamefont
  {F.}~\bibnamefont {Sette}},\ and\ \bibinfo {author} {\bibfnamefont
  {G.}~\bibnamefont {Ruocco}},\ }\bibfield  {title} {\bibinfo {title}
  {High-frequency longitudinal and transverse dynamics in water},\ }\href
  {https://doi.org/10.1103/physreve.71.011501} {\bibfield  {journal} {\bibinfo
  {journal} {Phys. Rev. E}\ }\textbf {\bibinfo {volume} {71}},\ \bibinfo
  {pages} {011501} (\bibinfo {year} {2005})}\BibitemShut {NoStop}%
\bibitem [{\citenamefont {Nosenko}\ \emph {et~al.}(2002)\citenamefont
  {Nosenko}, \citenamefont {Goree}, \citenamefont {Ma},\ and\ \citenamefont
  {Piel}}]{NosenkoPRL2002}%
  \BibitemOpen
  \bibfield  {author} {\bibinfo {author} {\bibfnamefont {V.}~\bibnamefont
  {Nosenko}}, \bibinfo {author} {\bibfnamefont {J.}~\bibnamefont {Goree}},
  \bibinfo {author} {\bibfnamefont {Z.~W.}\ \bibnamefont {Ma}},\ and\ \bibinfo
  {author} {\bibfnamefont {A.}~\bibnamefont {Piel}},\ }\bibfield  {title}
  {\bibinfo {title} {Observation of shear-wave {M}ach cones in a {2D}
  dusty-plasma crystal},\ }\href
  {https://doi.org/10.1103/physrevlett.88.135001} {\bibfield  {journal}
  {\bibinfo  {journal} {Phys. Rev. Lett.}\ }\textbf {\bibinfo {volume} {88}},\
  \bibinfo {pages} {135001} (\bibinfo {year} {2002})}\BibitemShut {NoStop}%
\bibitem [{\citenamefont {Nosenko}\ \emph {et~al.}(2003)\citenamefont
  {Nosenko}, \citenamefont {Goree}, \citenamefont {Ma}, \citenamefont {Dubin},\
  and\ \citenamefont {Piel}}]{NosenkoPRE2003}%
  \BibitemOpen
  \bibfield  {author} {\bibinfo {author} {\bibfnamefont {V.}~\bibnamefont
  {Nosenko}}, \bibinfo {author} {\bibfnamefont {J.}~\bibnamefont {Goree}},
  \bibinfo {author} {\bibfnamefont {Z.~W.}\ \bibnamefont {Ma}}, \bibinfo
  {author} {\bibfnamefont {D.~H.~E.}\ \bibnamefont {Dubin}},\ and\ \bibinfo
  {author} {\bibfnamefont {A.}~\bibnamefont {Piel}},\ }\bibfield  {title}
  {\bibinfo {title} {Compressional and shear wakes in a two-dimensional dusty
  plasma crystal},\ }\href {https://doi.org/10.1103/physreve.68.056409}
  {\bibfield  {journal} {\bibinfo  {journal} {Phys. Rev. E}\ }\textbf {\bibinfo
  {volume} {68}},\ \bibinfo {pages} {056409} (\bibinfo {year}
  {2003})}\BibitemShut {NoStop}%
\bibitem [{\citenamefont {Khrapak}\ and\ \citenamefont
  {Khrapak}(2022)}]{KhrapakJCP2022_1}%
  \BibitemOpen
  \bibfield  {author} {\bibinfo {author} {\bibfnamefont {S.~A.}\ \bibnamefont
  {Khrapak}}\ and\ \bibinfo {author} {\bibfnamefont {A.~G.}\ \bibnamefont
  {Khrapak}},\ }\bibfield  {title} {\bibinfo {title} {Freezing density scaling
  of fluid transport properties: Application to liquefied noble gases},\ }\href
  {https://doi.org/10.1063/5.0096947} {\bibfield  {journal} {\bibinfo
  {journal} {J. Chem. Phys.}\ }\textbf {\bibinfo {volume} {157}},\ \bibinfo
  {pages} {014501} (\bibinfo {year} {2022})}\BibitemShut {NoStop}%
\bibitem [{\citenamefont {Khrapak}\ and\ \citenamefont
  {Khrapak}(2023{\natexlab{b}})}]{KhrapakPoF2023}%
  \BibitemOpen
  \bibfield  {author} {\bibinfo {author} {\bibfnamefont {S.~A.}\ \bibnamefont
  {Khrapak}}\ and\ \bibinfo {author} {\bibfnamefont {A.~G.}\ \bibnamefont
  {Khrapak}},\ }\bibfield  {title} {\bibinfo {title} {Sound velocities in
  liquids near freezing: {D}ependence on the interaction potential and
  correlations with thermal conductivity},\ }\href
  {https://doi.org/10.1063/5.0157945} {\bibfield  {journal} {\bibinfo
  {journal} {Phys. Fluids}\ }\textbf {\bibinfo {volume} {35}},\ \bibinfo
  {pages} {077129} (\bibinfo {year} {2023}{\natexlab{b}})}\BibitemShut
  {NoStop}%
\bibitem [{\citenamefont {Khrapak}(2020{\natexlab{b}})}]{KhrapakPRR2020}%
  \BibitemOpen
  \bibfield  {author} {\bibinfo {author} {\bibfnamefont {S.~A.}\ \bibnamefont
  {Khrapak}},\ }\bibfield  {title} {\bibinfo {title} {Lindemann melting
  criterion in two dimensions},\ }\href
  {https://doi.org/10.1103/physrevresearch.2.012040} {\bibfield  {journal}
  {\bibinfo  {journal} {Phys. Rev. Research}\ }\textbf {\bibinfo {volume}
  {2}},\ \bibinfo {pages} {012040} (\bibinfo {year}
  {2020}{\natexlab{b}})}\BibitemShut {NoStop}%
\bibitem [{\citenamefont {Pedersen}\ \emph {et~al.}(2008)\citenamefont
  {Pedersen}, \citenamefont {Bailey}, \citenamefont {Schr{\o}der},\ and\
  \citenamefont {Dyre}}]{PedersenPRL2008}%
  \BibitemOpen
  \bibfield  {author} {\bibinfo {author} {\bibfnamefont {U.~R.}\ \bibnamefont
  {Pedersen}}, \bibinfo {author} {\bibfnamefont {N.~P.}\ \bibnamefont
  {Bailey}}, \bibinfo {author} {\bibfnamefont {T.~B.}\ \bibnamefont
  {Schr{\o}der}},\ and\ \bibinfo {author} {\bibfnamefont {J.~C.}\ \bibnamefont
  {Dyre}},\ }\bibfield  {title} {\bibinfo {title} {Strong pressure-energy
  correlations in van der {W}aals liquids},\ }\href
  {https://doi.org/10.1103/physrevlett.100.015701} {\bibfield  {journal}
  {\bibinfo  {journal} {Phys. Rev. Lett.}\ }\textbf {\bibinfo {volume} {100}},\
  \bibinfo {pages} {015701} (\bibinfo {year} {2008})}\BibitemShut {NoStop}%
\bibitem [{\citenamefont {Brazhkin}\ \emph
  {et~al.}(2012{\natexlab{a}})\citenamefont {Brazhkin}, \citenamefont {Fomin},
  \citenamefont {Lyapin}, \citenamefont {Ryzhov},\ and\ \citenamefont
  {Trachenko}}]{BrazhkinPRE2012}%
  \BibitemOpen
  \bibfield  {author} {\bibinfo {author} {\bibfnamefont {V.~V.}\ \bibnamefont
  {Brazhkin}}, \bibinfo {author} {\bibfnamefont {Y.~D.}\ \bibnamefont {Fomin}},
  \bibinfo {author} {\bibfnamefont {A.~G.}\ \bibnamefont {Lyapin}}, \bibinfo
  {author} {\bibfnamefont {V.~N.}\ \bibnamefont {Ryzhov}},\ and\ \bibinfo
  {author} {\bibfnamefont {K.}~\bibnamefont {Trachenko}},\ }\bibfield  {title}
  {\bibinfo {title} {Two liquid states of matter: A dynamic line on a phase
  diagram},\ }\href {https://doi.org/10.1103/physreve.85.031203} {\bibfield
  {journal} {\bibinfo  {journal} {Phys. Rev. E}\ }\textbf {\bibinfo {volume}
  {85}},\ \bibinfo {pages} {031203} (\bibinfo {year}
  {2012}{\natexlab{a}})}\BibitemShut {NoStop}%
\bibitem [{\citenamefont {Brazhkin}\ \emph {et~al.}(2013)\citenamefont
  {Brazhkin}, \citenamefont {Fomin}, \citenamefont {Lyapin}, \citenamefont
  {Ryzhov}, \citenamefont {Tsiok},\ and\ \citenamefont
  {Trachenko}}]{BrazhkinPRL2013}%
  \BibitemOpen
  \bibfield  {author} {\bibinfo {author} {\bibfnamefont {V.~V.}\ \bibnamefont
  {Brazhkin}}, \bibinfo {author} {\bibfnamefont {Y.~D.}\ \bibnamefont {Fomin}},
  \bibinfo {author} {\bibfnamefont {A.~G.}\ \bibnamefont {Lyapin}}, \bibinfo
  {author} {\bibfnamefont {V.~N.}\ \bibnamefont {Ryzhov}}, \bibinfo {author}
  {\bibfnamefont {E.~N.}\ \bibnamefont {Tsiok}},\ and\ \bibinfo {author}
  {\bibfnamefont {K.}~\bibnamefont {Trachenko}},\ }\bibfield  {title} {\bibinfo
  {title} {Liquid-gas'' transition in the supercritical region: fundamental
  changes in the particle dynamics},\ }\href@noop {} {\bibfield  {journal}
  {\bibinfo  {journal} {Phys. Rev. Lett.}\ }\textbf {\bibinfo {volume} {111}},\
  \bibinfo {pages} {145901} (\bibinfo {year} {2013})}\BibitemShut {NoStop}%
\bibitem [{\citenamefont {Brazhkin}\ \emph
  {et~al.}(2012{\natexlab{b}})\citenamefont {Brazhkin}, \citenamefont {Lyapin},
  \citenamefont {Ryzhov}, \citenamefont {Trachenko}, \citenamefont {Fomin},\
  and\ \citenamefont {Tsiok}}]{BrazhkinUFN2012}%
  \BibitemOpen
  \bibfield  {author} {\bibinfo {author} {\bibfnamefont {V.~V.}\ \bibnamefont
  {Brazhkin}}, \bibinfo {author} {\bibfnamefont {A.}~\bibnamefont {Lyapin}},
  \bibinfo {author} {\bibfnamefont {V.~N.}\ \bibnamefont {Ryzhov}}, \bibinfo
  {author} {\bibfnamefont {K.}~\bibnamefont {Trachenko}}, \bibinfo {author}
  {\bibfnamefont {Y.~D.}\ \bibnamefont {Fomin}},\ and\ \bibinfo {author}
  {\bibfnamefont {E.~N.}\ \bibnamefont {Tsiok}},\ }\bibfield  {title} {\bibinfo
  {title} {Where is the supercritical fluid on the phase diagram?},\ }\href
  {https://doi.org/10.3367/ufnr.0182.201211a.1137} {\bibfield  {journal}
  {\bibinfo  {journal} {Phys.-Usp.}\ }\textbf {\bibinfo {volume} {182}},\
  \bibinfo {pages} {1137} (\bibinfo {year} {2012}{\natexlab{b}})}\BibitemShut
  {NoStop}%
\bibitem [{\citenamefont {Khrapak}\ \emph {et~al.}(2025)\citenamefont
  {Khrapak}, \citenamefont {Formisano},\ and\ \citenamefont
  {Bove}}]{KhrapakPRE07_2025}%
  \BibitemOpen
  \bibfield  {author} {\bibinfo {author} {\bibfnamefont {S.~A.}\ \bibnamefont
  {Khrapak}}, \bibinfo {author} {\bibfnamefont {F.}~\bibnamefont {Formisano}},\
  and\ \bibinfo {author} {\bibfnamefont {L.~E.}\ \bibnamefont {Bove}},\
  }\bibfield  {title} {\bibinfo {title} {Transport properties of supercritical
  methane},\ }\href {https://doi.org/10.1103/kpsb-wvxl} {\bibfield  {journal}
  {\bibinfo  {journal} {Phys. Rev. E}\ }\textbf {\bibinfo {volume} {112}},\
  \bibinfo {pages} {015422} (\bibinfo {year} {2025})}\BibitemShut {NoStop}%
\bibitem [{\citenamefont {Bell}\ \emph {et~al.}(2020)\citenamefont {Bell},
  \citenamefont {Galliero}, \citenamefont {Delage-Santacreu},\ and\
  \citenamefont {Costigliola}}]{BellJCP2020}%
  \BibitemOpen
  \bibfield  {author} {\bibinfo {author} {\bibfnamefont {I.~H.}\ \bibnamefont
  {Bell}}, \bibinfo {author} {\bibfnamefont {G.}~\bibnamefont {Galliero}},
  \bibinfo {author} {\bibfnamefont {S.}~\bibnamefont {Delage-Santacreu}},\ and\
  \bibinfo {author} {\bibfnamefont {L.}~\bibnamefont {Costigliola}},\
  }\bibfield  {title} {\bibinfo {title} {An entropy scaling demarcation of gas-
  and liquid-like fluid behaviors},\ }\href {https://doi.org/10.1063/1.5143854}
  {\bibfield  {journal} {\bibinfo  {journal} {J. Chem. Phys.}\ }\textbf
  {\bibinfo {volume} {152}},\ \bibinfo {pages} {191102} (\bibinfo {year}
  {2020})}\BibitemShut {NoStop}%
\bibitem [{\citenamefont {Khrapak}(2022)}]{KhrapakJCP2022}%
  \BibitemOpen
  \bibfield  {author} {\bibinfo {author} {\bibfnamefont {S.~A.}\ \bibnamefont
  {Khrapak}},\ }\bibfield  {title} {\bibinfo {title} {Gas-liquid crossover in
  the {L}ennard-{J}ones system},\ }\href {https://doi.org/10.1063/5.0085181}
  {\bibfield  {journal} {\bibinfo  {journal} {J. Chem. Phys.}\ }\textbf
  {\bibinfo {volume} {156}},\ \bibinfo {pages} {116101} (\bibinfo {year}
  {2022})}\BibitemShut {NoStop}%
\end{thebibliography}%

\end{document}